\documentclass[]{aa} 
\usepackage[utf8]{inputenc}
\usepackage{amssymb}
\usepackage{amsmath}
\usepackage{graphicx} 
\usepackage{enumitem}
\usepackage{url}
\usepackage{times}
\setlist{nosep} %
\usepackage{sectsty}
\usepackage[colorlinks=true,linkcolor=blue,citecolor=blue]{hyperref}

\sectionfont{\fontsize{12}{12}\selectfont}
\subsectionfont{\fontsize{11}{11}\selectfont}

\usepackage{color}
\usepackage{ulem}

\newcommand{\resp}[1]{#1}
\newcommand{\resptwo}[1]{#1}

\title{Constraints on the merging binary neutron star mass distribution and equation of state based on the incidence of jets in the population\thanks{The source code behind the manuscript and all the computations is publicly available at \url{https://}}}

\titlerunning{BNS mass distribution and EoS constraints based on the jet incidence}

\author{Om Sharan Salafia\inst{1,2,3}, Alberto Colombo\inst{1,2}, 
Francesco Gabrielli\inst{4} and Ilya Mandel\inst{5,6,7}}

\authorrunning{O.~S.~Salafia, A.~Colombo, F.~Gabrielli \& I.~Mandel}

\institute{
 Università degli Studi di Milano-Bicocca, Dip.\ di Fisica ``G.\ Occhialini'', piazza della Scienza 3, I-20126 Milano (MI), Italy
 \and INFN -- Sezione di Milano-Bicocca, piazza della Scienza 2, I-20126 Milano (MI), Italy
 \and INAF -- Osservatorio Astronomico di Brera, via E. Bianchi 46, I-23807 Merate (LC), Italy
 \and Scuola Internazionale Superiore di Studi Avanzati (SISSA), via Bonomea 265, I-34136 Trieste (TS), Italy
 \and Monash Centre for Astrophysics, School of Physics and Astronomy, Monash University, Clayton, Victoria 3800, Australia
 \and ARC Center of Excellence for Gravitational Wave Discovery -- OzGrav
 \and Institute of Gravitational Wave Astronomy and School of Physics and Astronomy, University of Birmingham, Birmingham, B15 2TT, United Kingdom
}

\date{Draft, \today}

\abstract{
A relativistic jet has been produced in the single well-localised binary neutron star (BNS) merger detected to date in gravitational waves (GWs), and the local rates of BNS mergers and short gamma-ray bursts are of the same order of magnitude.  This suggests that jet formation is not a rare outcome for BNS mergers, and we show that this intuition can be turned into a quantitative constraint: at least about $1/3$ of GW-detected BNS mergers, and at least about $1/5$ of all BNS mergers, should produce a successful jet (90\% credible level). Whether a jet is launched depends on the properties of the merger remnant and of the surrounding accretion disc, which in turn are a function of the progenitor binary masses and equation of state (EoS).  The \resp{incidence} of jets in the population therefore carries information about the binary component mass distribution and EoS. Under the assumption that a jet can only be produced by a black hole remnant surrounded by a non-negligible accretion disc, we show how the jet \resp{incidence} can be used to place a joint constraint on the space of BNS component mass distributions and EoS. The result points to a broad mass distribution, with particularly strong support for masses in the $1.3-1.6\,\mathrm{M_\odot}$ range. The constraints on the EoS are shallow, but we show how they will tighten as the knowledge on the jet \resp{incidence} improves. We also discuss how to extend the method to include future BNS \resp{mergers}, with possibly uncertain jet associations.
}

\begin{document}

\maketitle


\section{Introduction}

Binary neutron star (BNS) mergers have long been considered \citep{Eichler1989} as promising candidate progenitors of at least some gamma-ray bursts (GRBs). Evidence accumulated through the years pointing to the two observational GRB classes (`long' and `short', \citealt{Kouveliotou1993}) being physically linked to two different progenitors, with collapsars \citep{Woosley1993} being the progenitors of long GRBs and BNS mergers (and possibly also neutron star - black hole mergers, \citealt{Mochkovitch1993}) being those of short GRBs (SGRBs, \citealt{Berger2014,DAvanzo2015}). The confirmation of the collapsar progenitor scenario for long GRBs came with the association of GRB980425 with supernova  SN1998bw \citep{Galama1998}.  The clear evidence \citep{Abbot2017_GW170817_GRB170817A,Mooley2018,Ghirlanda2019} in support of the presence of a relativistic jet, observed under a relatively large viewing angle with respect to its axis, in association with the GW170817 BNS merger detected by the Advanced Laser Interferometer Gravitational-wave Observatory (aLIGO, \citealt{aLIGO2015}) and Virgo \citep{Acernese2014} network and the fact that the inferred characteristics of such jet indicate \citep[e.g.][]{Salafia2019} that an on-axis observer would have observed an emission consistent with previously known SGRBs, provided the long-awaited smoking gun of the BNS-SGRB progenitor scenario. Still, the question whether \textit{all} BNS mergers produce a jet (and whether other progenitors contribute a significant fraction of SGRBs) remains open. The answer to this question encodes information about the conditions that lead to the launch of a jet and to its successful propagation up to the location where the observable emission is produced.

The conditions for the launch of a relativistic jet in the immediate aftermath of a BNS merger are set by the jet-launching mechanism, which is not entirely settled: while the \citet{Blandford1977} mechanism -- which involves the extraction of energy from a spinning black hole by magnetic fields supported by an accretion disc -- seems the most favoured one, several authors \citep[e.g.][]{Dai1998,Zhang2001,Metzger2008,DallOsso2011,Bernardini2012a,Bernardini2012b,Bernardini2013,Rowlinson2013,Gompertz2013,Stratta2018,Strang2019,Sarin2020} invoked a rapidly spinning, highly magnetized neutron star (a proto-magnetar) to explain some observational features such as the so called `plateaux' \citep{Nousek2006} in early X-ray afterglows and the detection of `extended emission' \citep{Norris2006} in gamma-rays beyond the usual short-duration prompt emission.  On the other hand, many models of the early X-ray afterglow and extended emission do not require a proto-magnetar \citep[e.g.][]{Rees1998,Zhang2006,Genet2007,Yamazaki2009,Leventis2014,Duffell2015,Beniamini2020,Oganesyan2020,Ascenzi2020,Barkov2021,Duque2021arXiv}, and the production of a successful jet by such a compact object, while possible \citep{Usov1992,Thompson1994,Thompson2004,Metzger2011,Mosta2020}, is theoretically disfavoured (e.g.~\citealt{Ciolfi2020_nomagnetar}, see also \S\ref{sec:BZconditions}). 

Whether a BNS merger satisfies the jet-launching conditions, whatever they are, depends on the \resp{progenitor system parameters} and on the equation of state (EoS) of neutron star matter \citep{Fryer2015,Piro2017}. \resp{Indeed,} despite their complexity, the merger dynamics are mostly deterministic, so \resp{that} the properties of the post-merger remnant and any accretion disc are directly linked to the progenitor binary masses and, in principle, spins and magnetic fields. Except for extreme cases, however, pre-merger neutron star spins and magnetic fields are thought to have limited impact (e.g.~\citealt{East2019,Dudi2021,Papenfort2022} for spins; e.g.~\citealt{Giacomazzo2009,Lira2022} for magnetic fields). Moreover, the magnetic field in the merger remnant and accretion disc is thought to undergo amplification due to Kelvin-Helmholtz instabilities and dynamo processes \citep[e.g.][]{Kiuchi2014,Kiuchi2015,Palenzuela2021}, leading to the loss of memory about the initial magnetization.
Therefore, the fraction of jet-launching systems in the BNS merger population is largely determined by the distribution of the progenitor binary masses and by the properties of neutron star matter, that is, by the (not yet well-known) EoS of matter beyond the nuclear saturation density. The incidence of jets in the population therefore carries information about both binary stellar evolution and nuclear physics, which can be investigated especially through multi-messenger observations of these sources. 

In principle, a relativistic jet may be launched, but not be able to propagate through the cloud of merger ejecta and break out of it, and some authors suggested \citep[e.g.][]{Moharana2017} that this is a frequent outcome. On the other hand, the expected properties of BNS merger ejecta and those of SGRB jets (also in light of GW170817) suggest that the vast majority of jets are successful in breaking out \citep{Duffell2018,Beniamini2019,Salafia2020}. \resp{In this article, we therefore consider systems that satisfy our `jet-launching conditions' to always lead to a successful jet break out.}

\resp{In what follows,} we derive the posterior probability distributions on the \resp{jet incidence, i.e.\ the fraction of systems that launch a jet,} in BNS mergers based on the presence of a jet in GW170817 (\S\ref{sec:fjet_GW}) and on the comparison of local rate densities of SGRBs and BNS mergers (\S\ref{sec:fj_from_R0}). We then describe a framework to model the jet \resp{incidence} under the physically-motivated assumption that launching a jet in the aftermath of the merger requires the collapse of the remnant to a black hole on a short time scale and the presence of a non-negligible accretion disc (\S\ref{sec:Modelling_fjet}). Within this framework, we show that knowledge of the jet \resp{incidence} can be used to constrain the BNS mass distribution and the EoS (\S\ref{sec:derivation}). We show (\S\ref{sec:mass_distrib_constraints}) that, within this framework, the currently available information on the jet \resp{incidence} leads to interesting constraints on the mass distribution, while it does not lead to informative constraints on the EoS.  Nevertheless, in \S\ref{sec:EoS_constraints} we show illustrative examples of EoS constraints that can be placed in the future, when the jet \resp{incidence} will be known with better precision. In Appendix~\ref{sec:fjGW_multiple} we show how the methodology can be extended to include multiple events with possibly uncertain jet associations, and in  \S\ref{sec:discussion} we discuss our results and suggest how this methodology can be modified to become part of Bayesian hierarchical population studies of BNS mergers.

\section{Jet incidence from observations}
\label{sec:fjet_estimation}
\subsection{GW-detectable sub-population: binomial argument}\label{sec:fjet_GW}
Let us call $f_\mathrm{j}$ the fraction of binary neutron star (BNS) mergers in a population that produce a successful relativistic jet.  At the time of writing, the only binary neutron star merger detected in GWs whose localisation was sufficiently tight and close-by as to permit a thorough electromagnetic follow-up was GW170817, and a relativistic jet has been clearly detected \resp{in association to it} \citep{Abbot2017_GW170817_GRB170817A,Mooley2018,Ghirlanda2019}. \resp{As we show in Appendix \ref{sec:fjGW_multiple}, the poorly constrained sky localisation and larger distance of the other GW-detected BNS merger event \citep[GW190425, ][]{Abbott2020_GW190425} prevents a useful constraint on the presence or absence of a jet; this event therefore does not carry information about the jet incidence and we ignore it here.} Let us model the production of a jet in the GW-detectable BNS subpopulation as a binomial process with single-event success probability $f_\mathrm{j,GW}$. The probability that a particular set of $k$ events out of a total of $n$ mergers produce a successful jet is then given by \begin{equation}
 P(k\,|\,n,f_\mathrm{j,GW})=f_\mathrm{j,GW}^k(1-f_\mathrm{j,GW})^{n-k},
 \label{eq:fjGWlikelihood}
\end{equation}
where the usual binomial coefficient is not present because the events can be distinguished.
For a single successful event after a single observation,
this is clearly $P(1\,|\,1,f_\mathrm{j,GW})=f_\mathrm{j,GW}$. By Bayes' theorem, the posterior probability density of $f_\mathrm{j,GW}$ is
\begin{equation}
 P(f_\mathrm{j,GW}\,|\,1,1) \propto f_\mathrm{j,GW} \pi(f_\mathrm{j,GW})
 \label{eq:fjGWposterior}
\end{equation}
where $\pi(f_\mathrm{j,GW})$ is the prior probability density on $f_\mathrm{j,GW}$. An intuitive and widely adopted choice for an uninformative prior probability is a uniform distribution $\pi(f_\mathrm{j,GW})=\pi_\mathrm{U}=1$ over the domain of the parameter, which is $f_\mathrm{j,GW}\in [0,1]$ in our case. A possibly more desirable choice \resp{of an uninformative prior is one that is invariant under re-parametrization} \citep{Jeffreys1946}. In this particular case, \resp{this `Jeffreys' prior} reads $\pi(f_\mathrm{j,GW})=\pi_\mathrm{J}(f_\mathrm{j,GW})=f_\mathrm{j,GW}^{-1/2}(1-f_\mathrm{j,GW})^{-1/2}$. With the former choice, the cumulative posterior probability of $f_\mathrm{j,GW}$ is 
\begin{equation}
 C_\mathrm{U}(f_\mathrm{j,GW}\,|\,1,1)=f_\mathrm{j,GW}^2
\end{equation}
while for the latter choice it takes a slightly more complicated form, namely 
\begin{equation}
C_\mathrm{J}(f_\mathrm{j,GW}\,|\,1,1)=\frac{2}{\pi}\left[\arcsin\left(\sqrt{f_\mathrm{j,GW}}\right)-\sqrt{f_\mathrm{j,GW}-f_\mathrm{j,GW}^2}\right].
\end{equation}
\begin{figure}
 \includegraphics[width=\columnwidth]{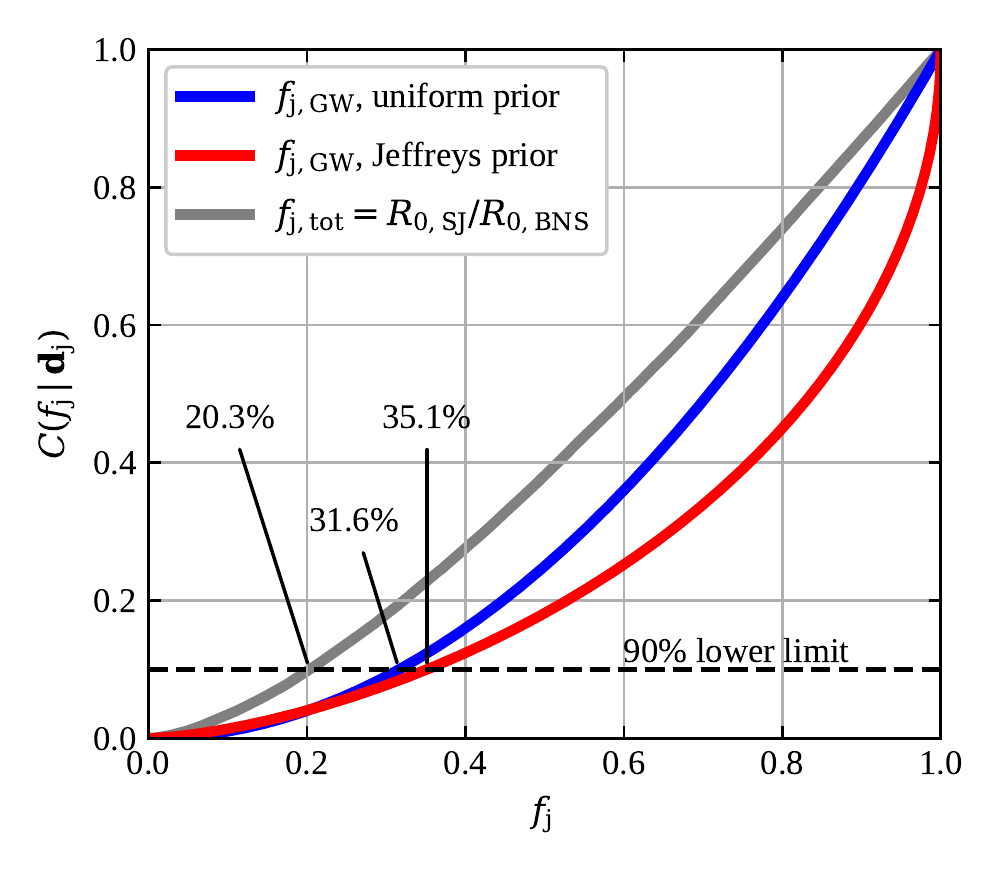}
 \caption{\small{Cumulative posterior probability of the BNS successful jet \resp{incidence}. The solid red (blue) line refers to the fraction $f_\mathrm{j,GW}$ of GW-detectable BNS mergers that produce a successful jet, based on having observed \resp{the} single successful event \resp{GW170817} and adopting the Jeffreys (uniform) prior. The solid grey line refers to the jet \resp{incidence} $f_\mathrm{j,tot}$ \resp{in the entire BNS merger population}, based on the ratio of the local rate \resp{of short GRB jets} $R_\mathrm{0,SJ}$ to the BNS rate $R_\mathrm{0,BNS}$. The implied 90\% credible lower limits are annotated.}}
 \label{fig:cumulative_posterior} 
\end{figure} 
From the cumulative posterior probability, a lower limit on $f_\mathrm{j,GW}$ at confidence level $\alpha$ can be derived by solving $C(f_\mathrm{j,GW}\,|\,1,1)=1-\alpha$ for $f_\mathrm{j,GW}$. In the uniform prior case, this can be done analytically, yielding $f_\mathrm{j,GW}\geq \sqrt{1-\alpha}$; for the Jeffreys prior the lower limit can be obtained numerically, or graphically from Figure~\ref{fig:cumulative_posterior}, which shows a plot of $C_\mathrm{U}$ and $C_\mathrm{J}$, with the implied 90\% confidence lower limit annotated. From this simple argument one can conclude, in agreement with\footnote{These works provide support for a large $f_\mathrm{j}$ based on the comparison of the single-event rate of GW170817 with the observed short gamma-ray burst rate.} \cite{Beniamini2019} and \cite{Ghirlanda2019}, that the relativistic jet in GW170817 implies that a large fraction -- at least about one third at the 90\% confidence level -- of GW-detectable BNS mergers should produce the same outcome, unless we have been extremely lucky in this first case. The corresponding 3-sigma lower limits are 5.2\% (3.4\%) for the uniform (Jeffreys) prior, showing that a fraction $f_\mathrm{j,GW}$ lower than several percent is highly unlikely. 
\label{sec:f_jet_min}

\subsection{Whole population: SGRB \textit{vs} BNS local rate}\label{sec:fj_from_R0}

\begin{figure}
 \centering
 \includegraphics[width=\columnwidth]{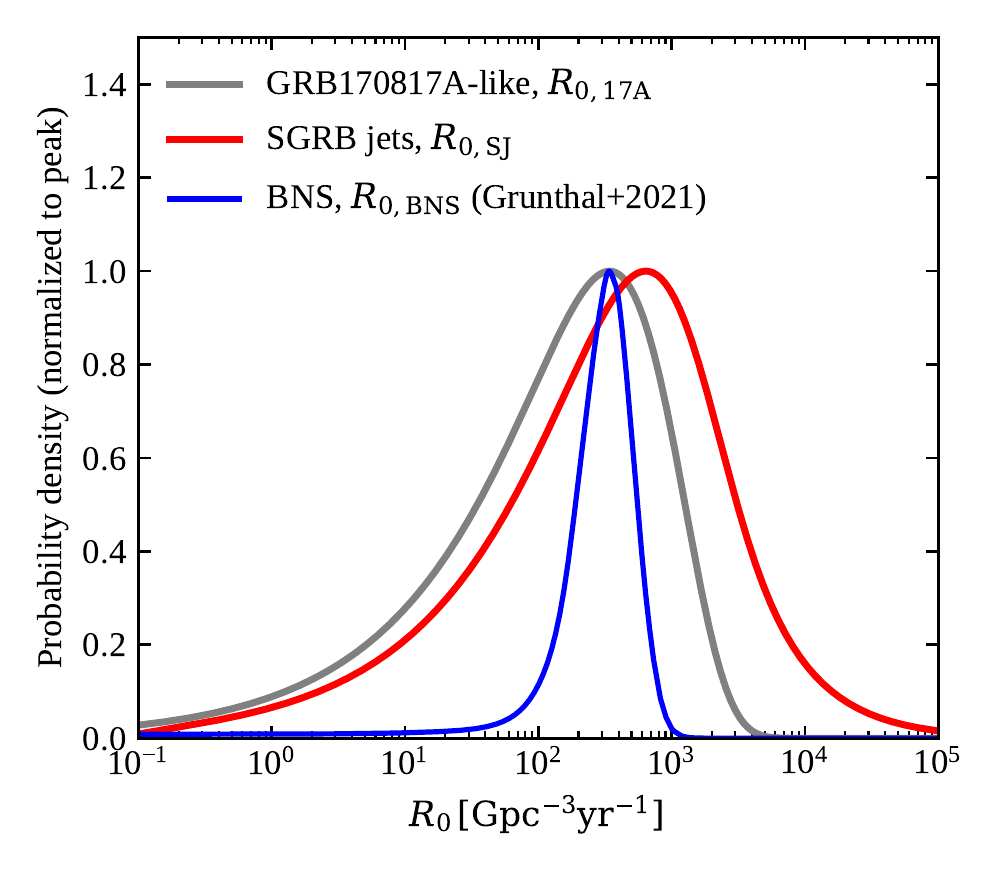}
 \caption{Posterior probability distribution on the local rate density of GRB170817A-like events (\resp{grey} solid line, Eq.~\ref{eq:P(R0)}), \resp{SGRB jets in general (regardless of orientation, red solid line, Eq.~\ref{eq:PR0SJ}), and} BNS mergers \resp{(as determined from observations of Galactic systems,} blue solid line, \citealt{Grunthal2021}).} 
 \label{fig:R0_R0BNS_comparison} 
\end{figure}

Another route to constraining $f_\mathrm{j}$ is that of comparing the local rate \resp{density} $R_\mathrm{0,SJ}$ of short GRB \resp{jets} \resp{(that is, the beaming-corrected SGRB rate density at $z=0$)} to that of BNS mergers, $R_\mathrm{0,BNS}$. \resp{The resulting jet incidence estimate} $f_\mathrm{j,tot}=R_\mathrm{0,SJ}/R_\mathrm{0,BNS}$ \resp{then} applies to all binaries (in contrast with that obtained in the previous section, which applies to the subpopulation of GW-detectable binaries). In a recent manuscript \citep{LVC2021_GWTC3pop} describing compact binary merger population analyses including data from the GWTC-3 catalog \citep{LVC2021_GWTC3}, the LIGO-Virgo-KAGRA (LVK) Collaboration found local BNS rates in the range $R_\mathrm{0,BNS}\in [10-1700]\,\mathrm{Gpc^{-3}\,yr^{-1}}$ (union of the 90\% credible intervals obtained through different analyses), the broad range being partly due to the highly uncertain mass distribution. An independent estimate, based on a larger sample of events, can be made based on known Galactic double neutron stars \citep{Kim2003}. A recent study (\citealt{Grunthal2021}; see also \citealt{Pol2020}), which includes observational insights on the beam shape and viewing geometry of the PSRJ1906+0746 pulsar (which contributes significantly to the total rate), finds a Milky Way BNS merger rate $\mathcal{R}_\mathrm{MW}=32_{-9}^{+19}\,\mathrm{Myr^{-1}}$ (\resp{maximum \textit{a posteriori} and 90\% credible range}). Assuming a Milky-Way-Equivalent galaxy density $\rho_\mathrm{MWEG}=0.0116\pm 0.0035\,\mathrm{Mpc^{-3}}$ in the local Universe (\citealt{Abadie2010,Kopparapu2008}, \resp{assuming a one-sigma local star formation rate uncertainty of $\sim 30\%$, \citealt{MadauFragos2017})}, this translates to $R_\mathrm{0,BNS}=390_{-256}^{+303}\,\mathrm{Gpc^{-3}\,yr^{-1}}$ (median and 90\% credible range, blue line in Fig.~\ref{fig:R0_R0BNS_comparison}). In what follows, we will adopt this BNS merger rate estimate as our reference \resp{value}\footnote{Double neutron star binaries containing an observable pulsar clearly represent only a subpopulation of merging binaries. While \citet{Grunthal2021} carefully account for selection effects and for the pulsar lifetimes in \resp{inferring the total number and merger rate of Galactic double neutron stars from the observed population}, they do not account for a possible fraction of the population that is not currently detected by radio surveys, which may form through different channels \citep[e.g.][]{VignaGomez2021} and may account for massive systems such as GW190425 \citep{Abbott2020_GW190425}. Such systems, on the other hand, are unlikely to constitute a dominant fraction of the total (as also suggested by the relative rate densities of GW170817 \resp{and} GW190425), and thus the systematic error that stems from using the \citet{Grunthal2021} rate as a proxy for the total BNS rate is most likely less than a factor of 2.  \resp{The uncertainty in the conversion between the Galactic and local volumetric merger rates incorporated into our analysis may therefore underestimate the total merger rate uncertainty.}}.

The local SGRB \resp{jet} rate density is poorly constrained due to the inherent difficulty in disentangling the luminosity function and redshift distribution of the cosmological \resp{SGRB} population, given the current low number of events with a measured redshift, the complexity of selection effects, and the fact that it is essentially impossible to constrain the low-end of the luminosity function (roughly below $5\times 10^{49}\,\mathrm{erg/s}$). This emerges clearly from the diversity of local rate values present in the literature \citep{Abbott2021_SGRBsearch,Mandel2021rates,Tan2020,Liu2019,Sun2017, Ghirlanda2016,Wanderman2015,Coward2012}. \resp{The uncertainty on the beaming-corrected rate density} is exacerbated by the difficulty in determining the average beaming factor (and its likely dependence on distance) from the available data. Nevertheless, \resp{here we show that a robust} lower limit on $R_\mathrm{0,SJ}$ \resp{can be} set \resp{based on} the single-event rate of GRB170817A, \resp{as follows. Let $V_\mathrm{eff}$ be the effective comoving volume over which \textit{Fermi/GBM} is sensitive to an event with the same intrinsic properties as GRB178017A. The rate density $R_{0,\mathrm{17A}}$ that can be derived conservatively assuming GBM to have observed a single SGRB jet in that volume over its entire time of operation $T=13\,\mathrm{yr}$, neglecting any beaming factor, is then a strict lower limit to $R_\mathrm{0,SJ}$.} 
By carefully modelling the GBM sensitive volume \resptwo{based on the peak fluxes of observed short GRBs} (as described in Appendix~\ref{sec:local_sgrb_rate}) we obtain \resp{the posterior distribution on $R_\mathrm{0,17A}$ shown by the red line in Fig.~\ref{fig:R0_R0BNS_comparison}, whose median and 90\% credible interval are} $R_\mathrm{0,17A}=342_{-337}^{+1800}\,\mathrm{Gpc^{-3}\,yr^{-1}}$ \resptwo{(in agreement with e.g.\ \citealt{DellaValle2018})}. Given this piece of information, 
our knowledge of $R_\mathrm{0,SJ}$ \resp{can be modelled as}
\begin{equation}
\begin{split}
 & P(R_\mathrm{0,SJ}\,|\,R_\mathrm{0,17A})\propto \\
 & \pi(R_\mathrm{0,SJ})\int \Theta(R_\mathrm{0,SJ}-R_\mathrm{0,17A})\pi(R_\mathrm{0,17A})\,\mathrm{dR_\mathrm{0,17A}},
\end{split}
\label{eq:PR0SJ}
\end{equation}
where $\Theta$ is the Heaviside step function, that is 
\begin{equation}
 \Theta(x) = \left\lbrace\begin{array}{lr}
                          0 & x<0\\
                          1 & x\geq 0
                         \end{array}\right..
\end{equation}
We use the \resp{above mentioned} $R_\mathrm{0,17A}$ posterior as our prior $\pi(R_\mathrm{0,17A})$, and \resp{adopt} a log-uniform \resp{prior} $\pi(R_\mathrm{0,SJ})\propto R_\mathrm{0,SJ}^{-1}$ \resp{on the SGRB jet rate density, which is a conservative choice as it favours values closer to the lower limit $R_\mathrm{0,17A}$, and hence lower values of $f_\mathrm{j,tot}$. The result is shown in Fig.~\ref{fig:R0_R0BNS_comparison}, along with the posterior on $R_\mathrm{0,17A}$ and that on $R_\mathrm{0,BNS}$ based on \citet{Grunthal2021}.}

\begin{figure} 
 \centering
 \includegraphics[width=\columnwidth]{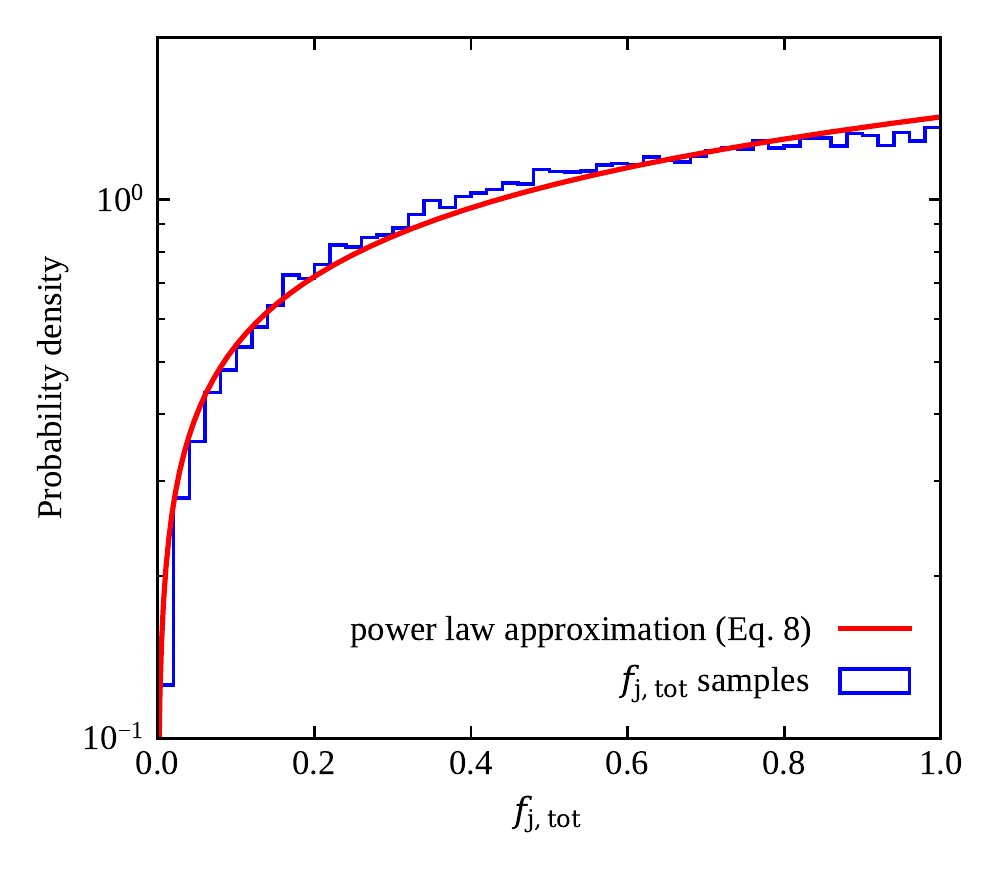}
 \caption{Posterior distribution on the fraction $f_\mathrm{j,tot}$ of BNS mergers accompanied by jets, computed \resp{as described in \S\ref{sec:fj_from_R0}}. The blue histogram shows the distribution of samples constructed as described in \resp{the text, that approximate Eq.~\ref{eq:fjtot_posterior}}, while the red line shows the analytical approximation $P(f_\mathrm{j,tot}\,|\,\vec d_\mathrm{tot})\propto f_\mathrm{j,tot}^\gamma$ with $\gamma=0.42$.}
 \label{fig:fj_posterior_R0_R0BNS_comparison} 
\end{figure}

\resp{Finally, we} compute the $f_\mathrm{j,tot}$ posterior distribution as the ratio distribution of the $R_\mathrm{0,SJ}$ and $R_\mathrm{0,BNS}$ distributions, \resp{with the additional constraint $f_\mathrm{j,tot}\leq 1$, namely}
\begin{equation}\label{eq:fjtot_posterior}
\begin{split}
 & P(f_\mathrm{j,tot}\,|\,\vec d_\mathrm{tot})\propto \iint \mathrm{d}R_\mathrm{0,SJ} \, \mathrm{d}R_\mathrm{0,BNS} \,\delta\left(f_\mathrm{j,tot}-\frac{R_\mathrm{0,SJ}}{R_\mathrm{0,BNS}}\right)\times \\
 & \times\Theta(R_\mathrm{0,BNS}-R_\mathrm{0,SJ})  P(R_\mathrm{0,SJ}\,|\,R_\mathrm{0,17A})\pi(R_\mathrm{0,BNS}),
\end{split}
\end{equation}
\resp{where $\delta$ is the Dirac delta function and $\vec d_\mathrm{tot}$ represents the data which inform our estimates of $R_\mathrm{0,SJ}$ and $R_\mathrm{0,BNS}$ and, hence, the above posterior on $f_\mathrm{j,tot}$. }
\resp{To evaluate Eq.~\ref{eq:fjtot_posterior}} in practice, we produce $\lbrace f_{\mathrm{j,tot},i} \rbrace_{i=1}^{N}$ samples by the following procedure: (1) we extract an $R_{\mathrm{0,SJ},i}$ sample from a log-uniform distribution in the range $10^{-1}-10^4\,\mathrm{Gpc^{-3}\,yr^{-1}}$ \resp{(that is, a broad enough range as to cover the support of both $R_\mathrm{0,17A}$ and $R_\mathrm{0,BNS}$)}, and an $R_{\mathrm{0,17A},i}$ sample from the red distribution in Fig.~\ref{fig:R0_R0BNS_comparison}; (2) if $R_{\mathrm{0,SJ},i}<R_{\mathrm{0,17A},i}$ we go back to (1); (3) we extract an $R_{\mathrm{0,BNS},i}$ sample from the blue distribution in Fig.~\ref{fig:R0_R0BNS_comparison}; (4) if $f_{\mathrm{j,tot},i}=R_{\mathrm{0,SJ},i}/R_{\mathrm{0,BNS},i}>1$ we reject the sample and go back to (1), otherwise we store it. The resulting cumulative $f_\mathrm{j,tot}$ distribution is shown by the grey line in Figure~\ref{fig:cumulative_posterior}, while two approximations of the probability density are shown in Fig.~\ref{fig:fj_posterior_R0_R0BNS_comparison}.  The implied 90\% credible lower limit on $f_\mathrm{j}$ is 21\% (2.4\% at $3\sigma$). As shown in Fig.~\ref{fig:fj_posterior_R0_R0BNS_comparison}, the associated probability density is well approximated by a power law 
\begin{equation}
 P(f_\mathrm{j,tot}\,|\,\vec d_\mathrm{tot})\propto f_\mathrm{j,tot}^{\gamma},
 \label{eq:P(fj from R0)}
\end{equation}
with a best fit value $\gamma=0.42$.
%
  
\resp{We neglect} the potential contribution of neutron star - black hole mergers \resp{to the beaming-corrected local rate density of SGRB jets.  This is justified by the fact that} the intrinsic BNS and neutron star - black hole rates are comparable -- \citealt{Abbott2021_NSBH} -- and most likely only a small fraction of the latter result in the tidal disruption of the neutron star, \resp{given the expected dearth of rapidly spinning BHs in these systems} (see e.g.~\citealt{Broekgaarden2021_NSBH} and \citealt{Zappa2019}). Other works take the alternative route of constraining the beaming factor (or the angular dependence of the jet emission properties) from the comparison \resp{of the observed rates of BNS mergers and SGRBs} \citep[e.g.][]{Williams2018,Biscoveanu2020,Farah2020,Hayes2020}, or of estimating the future joint BNS-SGRB detection rates as a function of such factor \citep[e.g.][]{Chen2013,Clark2015}.

\section{Modelling the jet incidence}
\label{sec:Modelling_fjet}
\subsection{Blandford-Znajek jet launching conditions}
\label{sec:BZconditions}

The gamma-ray burst jet launching mechanism is still under debate (see \citealt{Salafia2021} for a recent discussion \resp{and \citealt{Kumar2015} for a review}). In the case of a binary neutron star merger progenitor, the possible mechanisms are restricted to those compatible with the characteristics of the merger remnant: depending on the component neutron star masses and equation of state, the merger can lead either to the prompt formation (i.e.~happening on a dynamical timescale) of a black hole (BH) or to a proto-neutron star \citep{Burrows1986}. If the mass $M_\mathrm{rem}$ of the latter is below the maximum mass $M_\mathrm{TOV}$ that can be supported against collapse in a non-rotating neutron star, it will evolve to an indefinitely stable neutron star (NS); if the mass is above $M_\mathrm{TOV}$, but below the mass $M_\mathrm{max,rot}$ that can be supported by uniform rotation at the mass-shedding limit \citep{Goussard1997}, the remnant is said to be a 'supra-massive' neutron star (SMNS) and it can survive until electromagnetic spin-down eventually leads to collapse to a BH; if $M_\mathrm{rem}>M_\mathrm{max,rot}$, the proto-neutron star can still be supported \citep{Goussard1998} for a short time (typically $\lesssim 100\,\mathrm{ms}$) by differential rotation before collapsing, in which case it is termed a `hyper-massive' neutron star (HMNS).

Jets launched by neutron stars are observed in our Galaxy \citep[e.g.][]{Pavan2014,VanDenEijnden2018} and several authors have argued that a rapidly spinning, highly magnetized proto-neutron star merger remnant may be able to launch a relativistic jet \citep[e.g.][]{Usov1992,Thompson1994,Thompson2004,Metzger2011,Mosta2020}. On the other hand, the neutrino-driven \citep{Dessart2009,Perego2014} and magnetically-driven \citep{Ciolfi2020_winds} winds produced by a proto-neutron star during its early evolution are likely to load the surrounding environment with too many baryons, preventing a putative jet from reaching relativistic speeds \citep[e.g.][]{Ciolfi2020_nomagnetar}.

If the remnant is a BH (or it collapses to a BH in a time much shorter than the accretion time scale), following the amplification of the magnetic field by Kelvin-Helmholtz instabilities triggered at merger \citep{Obergaulinger2010,Kiuchi2015,Aguilera-Miret2020} \resp{and the development of a large-scale ordered magnetic field configuration with a non-negligible poloidal component}, the Blandford-Znajek jet-launching mechanism \citep{Blandford1977,Komissarov2001,Tchekhovskoy2010} can operate, possibly enhanced by energy deposition from the annihilation of neutrino-antineutrino pairs emitted by the inner accretion disc (\citealt{Eichler1989} -- even though this cannot be the dominant source of jet power, as found by \citealt{Just2016}).

Given the difficulties with the proto-neutron star central engine, in this work we assume that only a BH remnant (possibly, but not necessarily, formed after a HMNS phase) can launch a relativistic jet. Very broadly speaking, the conditions for the Blandford-Znajek mechanism to operate are the presence of a spinning black hole (BH), \resp{an accretion disc with a large-scale ordered magnetic field}, and a low-density funnel above the BH (along its rotation axis). Given the particular accretion conditions in the BNS post-merger phase and the expected predominantly toroidal configuration of the magnetic field in the disc \citep{Kiuchi2014,Kawamura2016}, the accretion-to-jet energy conversion efficiency $\eta = E_\mathrm{jet}/(M_\mathrm{disc}c^2)$ of the Blandford-Znajek mechanism in these systems \citep{Christie2019} is likely quite low, $\eta\sim 10^{-3}$, as found in numerical relativity simulations \citep{Ruiz2018} and in agreement with the estimate that can be made based on GW170817 \citep{Salafia2021}. Assuming this efficiency, in order to produce a (short) gamma-ray burst jet with a \resp{total} energy in the typical \citep{Fong2015} range $E_\mathrm{jet}\sim 10^{49}-10^{50}\,\mathrm{erg}$, a disc mass of about $M_\mathrm{disc}\sim 0.01-0.1\,\mathrm{M_\odot}$ is needed. 

\begin{figure}
 \centering
 \includegraphics[width=\columnwidth]{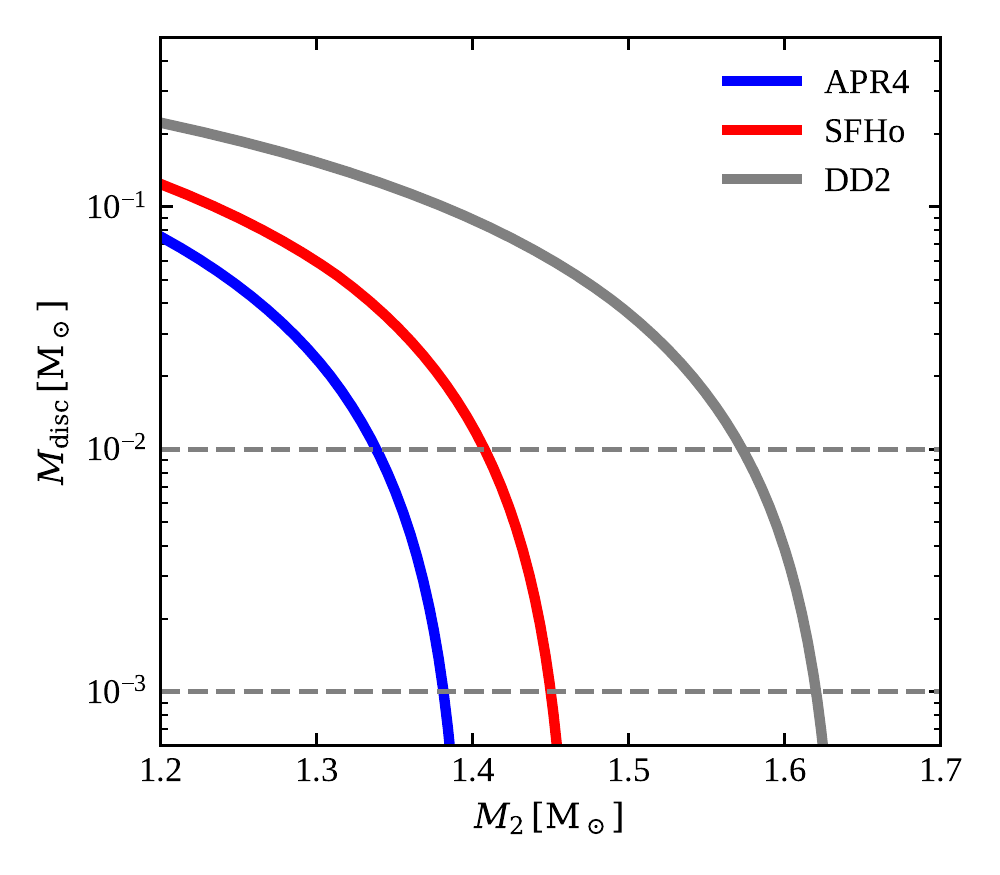}
 \caption{Dependence of the disc mass on the mass of the secondary component in the binary, according to the fitting formula by \cite{Kruger2020}. Different colors refer to different assumed equations of state, as shown in the legend.}
 \label{fig:mdisc_m2_dependence}
\end{figure}

Based on these arguments, we assume GRB jet launching to be possible only for HMNS or prompt BH remnants\footnote{We note that the inclusion of prompt BH collapse cases is justified by the fact that sufficiently asymmetric systems can produce massive accretion discs by the tidal disruption of the secondary neutron star, regardless of the nature of the remnant \citep{Bernuzzi2020}.}, provided that the accretion disc mass is, conservatively\footnote{We note that changing this to a more stringent value, e.g.~$M_\mathrm{disc,min}=10^{-2}\,\mathrm{M_\odot}$, would have a very limited impact on our results due to the steep dependence of the disc mass on the gravitational mass $M_2$ of the secondary component in the binary \citep{Kruger2020}, as demonstrated in Figure~\ref{fig:mdisc_m2_dependence}. For the same reason, the uncertainty in the disc mass fitting formula has little impact on our results.}, at least $M_\mathrm{disc,min}=10^{-3}\,\mathrm{M_\odot}$.  The first condition can be restated as the requirement that the gravitational mass $M_\mathrm{rem}$ of the merger remnant be at least as massive as the maximum mass that can be supported by uniform rotation $M_\mathrm{max,rot}=1.2 M_\mathrm{TOV}$, where $M_\mathrm{TOV}$ is \resp{set} by the chosen neutron star matter equation of state (EoS) and the factor 1.2 is essentially EoS-independent\footnote{A small deviation from this quasi-universal value is expected for equations of state with a first-order phase transition \citep[e.g.][]{Bozzola2019}.} \citep{Breu2016}. 

\subsection{Determination of the remnant type}\label{sec:remnant_determination}

In order to compute the remnant mass $M_\mathrm{rem}$, we invoke energy conservation for a merging binary with total mass $M=M_1+M_2$, \resp{equating the energy budget of the initial state (at formally infinite binary separation) to that of the final state after the remnant has formed}, in the form
\begin{equation}
 M c^2 = M_\mathrm{rem}c^2 + E_\mathrm{GW} +  E_\mathrm{disc} + E_\mathrm{ej} + E_\mathrm{\nu}
 \label{eq:binary_energy_cons}\, ,
\end{equation}
where $E_\mathrm{GW}$ is the energy radiated in gravitational waves, $E_\mathrm{disc}$ is the total energy associated to the accretion disc and $E_\mathrm{ej}$ is that associated to the ejecta (except for disc winds, which are included in $E_\mathrm{disc}$), and $E_\mathrm{\nu}$ is the energy lost in neutrinos by the remnant (in the absence of a prompt BH collapse). We compute $E_\mathrm{GW}$ using numerical-relativity-based fitting formulae from \cite{Zappa2018}, which include GW mass loss all the way from the inspiral to the early post-merger phase, and are implemented in the publicly available repository \texttt{bns\_lum} \citep{bns_lum}.  The disc energy in principle comprises its rest-mass, its gravitational potential energy, the energy associated to its rotation and \resp{that contained in the magnetic field}: the rotational and gravitational potential energy together amount approximately to $-G M_\mathrm{rem}M_\mathrm{disc}/2 R_\mathrm{disc} = - (R_\mathrm{g,rem}/R_\mathrm{disc})M_\mathrm{disc}c^2/2$ (where $R_\mathrm{g,rem}$ and $R_\mathrm{disc}$ are the remnant gravitational radius and the disc radius, respectively). Since very generally $R_\mathrm{disc}\gg R_\mathrm{g,rem}$, these terms can be safely neglected. \resp{As stated in the previous section, during the merger the magnetic field is amplified by Kelvin-Helmholtz instabilities, reaching typical energies of $\sim 10^{51}\,\mathrm{erg}$ \citep{Aguilera-Miret2020}. This is equivalent to less than $0.1\%$ of a solar mass, so that this term can also be neglected. Therefore, we keep} only the rest-mass contribution, that is $E_\mathrm{disc}\sim M_\mathrm{disc} c^2$. We compute the disc rest-mass $M_\mathrm{disc}$ using the fitting formula proposed in \citep[][\resp{their eq.~4, with their best-fit parameters, to which we refer hereafter as $M_\mathrm{disc,KF20}$}]{Kruger2020}. Similarly, since the typical neutron star merger ejecta velocities are subrelativistic, $v_\mathrm{ej}\sim 0.1-0.2\,c$, we neglect their kinetic energy and we write $E_\mathrm{ej}\sim M_\mathrm{ej}c^2$. The mass $M_\mathrm{ej}$ includes the dynamical ejecta, which we compute through the relevant fitting formula from \citep[][\resp{their eq.~6, with their best-fit coefficients}]{Kruger2020}, and  potentially also winds from a long-lived neutron star remnant. The typical mass of neutrino-driven winds from a HMNS is expected \citep{Dessart2009,Perego2014} to be on the order of $10^{-4}\,\mathrm{M_\odot}$ and can be safely neglected in our treatment; winds driven by magnetically-induced viscosity can be much more substantial, reaching $\sim 10^{-2}\,\mathrm{M_\odot}$ \citep{Ciolfi2020_winds} in sufficiently long-lived cases, but this is still well within the expected uncertainty of the fitting formula we employ for the disc mass \citep[][]{Kruger2020}, and therefore we do not include this component for simplicity. Similarly, since the peak neutrino luminosity of an HMNS remnant is expected to be below $L_\mathrm{\nu}<10^{53}\,\mathrm{erg/s}$ \citep{Dessart2009,Perego2014}, and the HMNS lifetime (\resp{which is of the order of} the \resp{viscous} timescale \resp{in the HMNS, since} differential rotation is damped \resp{by viscous angular momentum transport},  e.g.\ \citealt{Kiuchi2018}) is $t_\mathrm{HMNS}\lesssim 0.1\,\mathrm{s}$, the mass lost in neutrinos $M_\mathrm{\nu}=E_\mathrm{\nu}/c^2\sim L_\mathrm{\nu}t_\mathrm{HMNS}/c^2$ is well below one percent of a solar mass, and thus we set $E_\mathrm{\nu}=0$. 

\begin{figure*}
\centering
\includegraphics[width=\textwidth]{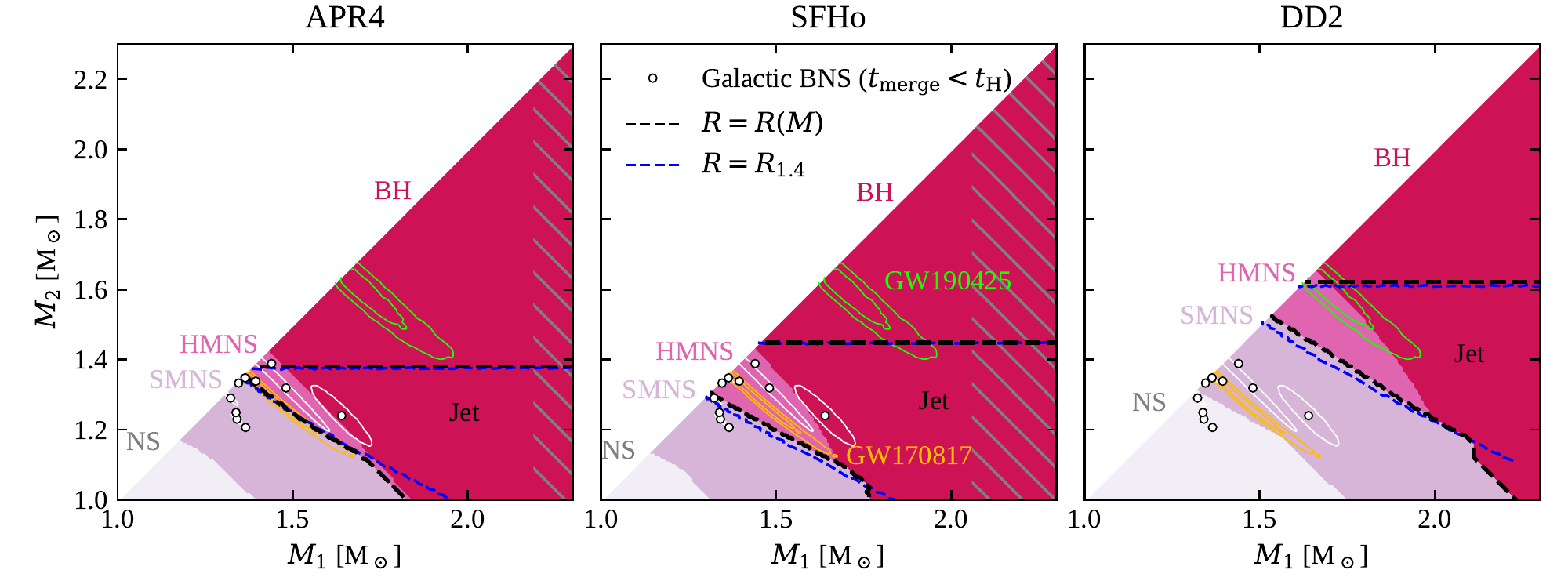}   
\caption{Merger remnant and GRB launching regions on the component mass plane. The filled coloured regions correspond to different BNS merger remnants on the $(M_1,M_2)$ plane, assuming different EoSs: APR4 (left-hand panel), SFHo (central panel) and DD2 (right-hand panel). Grey hatches mark masses that are above the maximum non-rotating mass supported by the EoS. The black dashed lines enclose the region where our GRB jet launching conditions are met, computed as described in \S\ref{sec:BZconditions} (the blue dashed lines, shown for comparison, are computed with the approximate method described in \S\ref{sec:approx_bzconditions}). White circles and contours show the best fit and 90\% credible contours of masses of known Galactic BNS systems \resp{whose coalescence time $t_\mathrm{merge}$ is shorter than} a Hubble time \citep[$t_\mathrm{H}$, data from ][]{Farrow2019}. 
The orange and green contours show the 50\% and 90\% credible regions for the component masses in GW170817 \citep{Abbott2019_GW170817_properties} and GW190425 \citep{Abbott2020_GW190425}, respectively, adopting low-spin priors.}
\label{fig:M1_M2_plane}
\end{figure*}

The employed fitting formulae depend on the neutron star masses and on their compactness or dimensionless tidal deformability: if an EoS is assumed, the neutron star radii can be computed from the mass-radius relation, and the tidal deformabilities from the `C-Love' universal relation \citep{Yagi2017}. This allows one to solve Eq.~\ref{eq:binary_energy_cons} for $M_\mathrm{rem}=M_\mathrm{rem}(M_1,M_2,\vec\theta_\mathrm{EoS})$, where $\vec\theta_\mathrm{EoS}$ represents a set of parameters that define the EoS. Setting $M_\mathrm{rem}<M_\mathrm{TOV}$ as the condition for an indefinitely stable NS remnant; $M_\mathrm{TOV}\leq M_\mathrm{rem}<1.2 M_\mathrm{TOV}$ as the condition for a SMNS remnant; $1.2 M_\mathrm{TOV}\leq M_\mathrm{rem}< M_\mathrm{thresh}$ as the condition for a HMNS remnant, and $M_\mathrm{rem}\geq M_\mathrm{thresh}$ for direct collapse to a BH, one can then determine the fate of the merger remnant based on the initial binary masses and on the EoS. Here $M_\mathrm{thresh}$ is the threshold mass for BH direct collapse, which can be computed employing the mass ratio-dependent fitting formula given in \citealt{Bauswein2021}. Alternative recent formulae exist (e.g.~\citealt{Tootle2021,Kashyap2021,Kolsch2021}), but we note that our results do not depend on this choice: the $M_\mathrm{thresh}$ line shown in Fig.~\ref{fig:M1_M2_plane} as the separation between the HMNS and BH remnant regions is only illustrative and it does not enter our assumed jet-launching conditions.

Figure \ref{fig:M1_M2_plane} shows the resulting regions that correspond to different expected remnants of BNS mergers on the $(M_1,M_2)$ plane, assuming three different EoSs, namely APR4 \citep[][]{Akmal1998}, SFHo and DD2 \citep[][]{Hempel2012}, which predict maximum masses $M_\mathrm{TOV}/\mathrm{M_\odot}=2.19$, 2.06 and 2.42, respectively (all of which are above the largest Galactic neutron star masses measured to date, \citealt{Lynch2013,Fonseca2021}), and radii in the range  $R_\mathrm{1.4}\in (11.30,13.25)\,\mathrm{km}$ for a $1.4\,\mathrm{M_\odot}$ neutron star, in agreement with constraints from GW170817 \citep{Abbott2018_GW170817_EoS,Abbott2019_GW170817_properties} and from NICER (\citealt{Miller2019,Raaijmakers2019,Raaijmakers2020,Landry2020}, \resp{with some tension in the case of DD2}). The black dashed lines enclose the region where our GRB jet launching conditions are met. The upper boundary is a horizontal line: this is due to the fact that the disc mass fitting formula by \cite{Kruger2020} depends only on the compactness of the least massive neutron star, but we verified \resp{that a weak dependence on the mass of the most massive NS is reflected in} other recent disc mass fitting formulae \citep{Dietrich2020,Barbieri2021}, \resp{which also lead to upper boundaries that are close to horizontal}. The blue dashed lines show the result obtained by adopting the simplified EoS dependence described in the next section, plugging in the $M_\mathrm{TOV}$ and $R_{1.4}$ values implied by each EoS. Confidence contours for the component masses of GW170817 \citep{Abbott2019_GW170817_properties} and GW190425 \citep{Abbott2020_GW190425} are shown for reference. For the two softer equations of state, APR4 and SFHo, our jet-launching conditions are most likely satisfied in GW170817 and also in a significant fraction of Galactic BNS (4/10 for APR4; 6/10 for SFHo). If the stiffer DD2 EoS is adopted, then GW170817 clearly falls in the supra-massive neutron remnant region: if such an EoS were found to best represent neutron star matter, then the jet-launching mechanism \resp{that operated in the case of GW170817} would have to be something different from Blandford-Znajek.

\subsection{Approximate two-parameter EoS dependence}
\label{sec:approx_bzconditions} 
Our assumed jet-launching conditions require, in practice, the merger remnant mass to satisfy $M_\mathrm{rem}>1.2 M_\mathrm{TOV}$ and the disc mass to satisfy $M_\mathrm{disc}>M_\mathrm{disc,min}$. A rough idea of where the former separation line stands can be obtained by writing a simplified remnant mass equation $M_\mathrm{rem}\sim M - E_\mathrm{GW}/c^2 - M_\mathrm{disc}$, thus neglecting the dynamical ejecta mass (in addition to all other terms which we already neglect for the reasons explained in the previous section). This leads to a critical total mass for jet launch $M_\mathrm{crit}\sim (1.2 M_\mathrm{TOV}+M_\mathrm{disc})/(1-\eta_\mathrm{GW})$, where $\eta_\mathrm{GW}=E_\mathrm{GW}/(Mc^2)$. This places the critical mass in the range $M_\mathrm{crit}/\mathrm
{M_\odot}\in [2.4,3.26]$ for $2\leq M_\mathrm{TOV}/\mathrm{M_\odot} \leq 2.5$, $M_\mathrm{disc}\leq 0.1 \mathrm{M_\odot}$ and $\eta_\mathrm{GW}\leq 0.05$. Assuming equal masses, the component masses are therefore in the range $M_{1,2}/\mathrm{M_\odot}\in [1.2,1.63]$. For most viable EoSs, the NS radius is approximately  constant within this mass range \citep[e.g.][]{Ozel2016}, so that one can quite safely assume it to be equal to $R_{1.4}$, i.e.~that of a reference $1.4\,\mathrm{M_\odot}$ NS. \resp{Fixing the NS radius to this value, the compactness of the secondary can be determined from its mass. The disc mass computed using the $M_\mathrm{disc,KF20}$ fitting formula thus becomes a function of $M_2$ only.} A reasonable estimate of $\eta_\mathrm{GW}$ can be obtained as the absolute value of the Keplerian orbital energy of two point masses $(M_1, M_2)$, at a $2R_{1.4}$ separation, divided by $M c^2$, which gives $\eta_\mathrm{GW}\sim \nu G M /4 R_{1.4} c^2 $, where $\nu=M_1M_2/M^2=(2+q+q^{-1})^{-1}$ is the symmetric mass ratio. This effectively reduces the EoS dependence of $M_\mathrm{crit}$ to only two parameters, namely $\vec\theta_\mathrm{EoS}=\lbrace R_{1.4},M_\mathrm{TOV}\rbrace$. In Figure~\ref{fig:M1_M2_plane} we show the $M_\mathrm{crit}$ obtained with this method (oblique blue dashed lines). To obtain also the line beyond which $M_\mathrm{disc}<10^{-3}\,\mathrm{M_\odot}$ (horizontal blue dashed lines), we employed the $M_\mathrm{disc,KF20}$ fitting formula keeping the NS radius fixed at $R_{1.4}$. These examples show that the approximate method yields results that differ from the more accurate ones (black dashed lines) at the percent level. \resp{In the remainder of this work, we adopt the simpified 2-parameter dependence described above.}

\subsection{EoS priors and jet probability on the $(M_1,M_2)$ plane}
\label{sec:EoS_priors}

\begin{figure}
 \centering
 \includegraphics[width=\columnwidth]{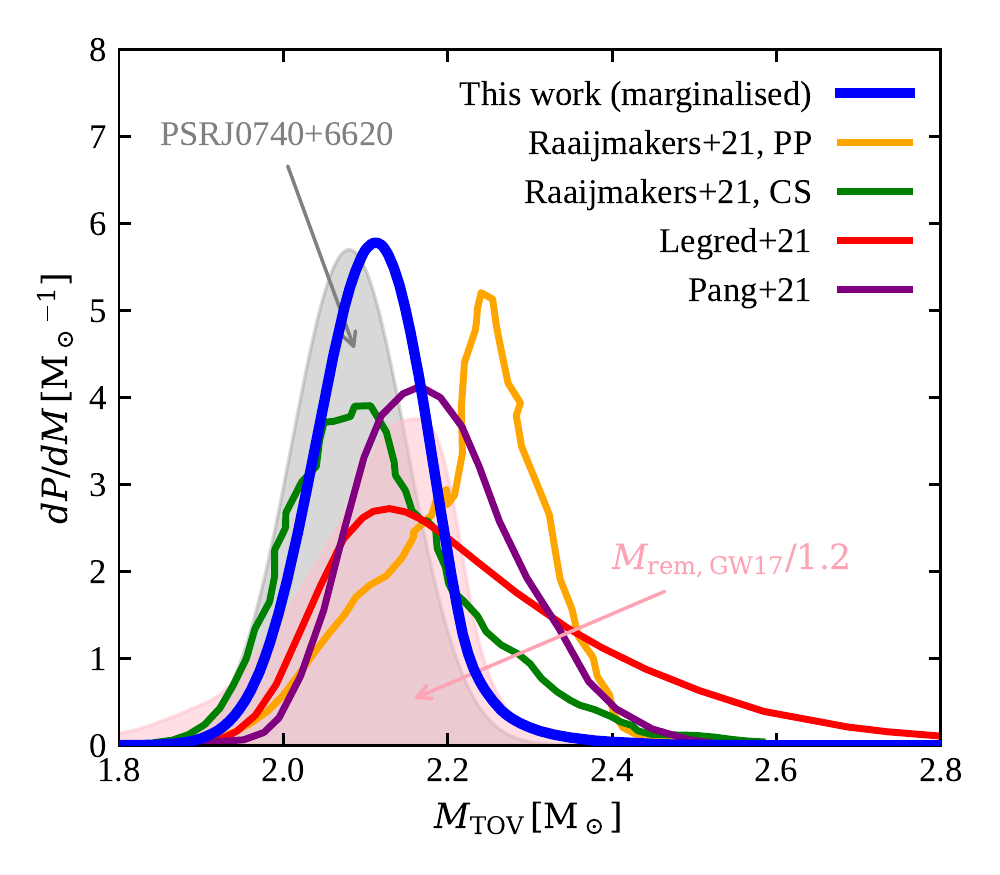} 
 \caption{Our prior on $M_\mathrm{TOV}$ compared to recent model-dependent constraints in the literature. The blue thick line shows the \resp{marginalised} $M_\mathrm{TOV}$ prior we adopt in this work, which is constructed using the latest PSRJ0740+6620 mass measurement \citep[][grey filled area]{Fonseca2021} as a lower limit and the \resp{remnant} mass of GW170817 divided by 1.2 (pink filled area) as an upper limit (see text for an explanation). \resp{The remaining lines show results that combine multi-messenger constraints within different frameworks: the orange and green lines show the results obtained by \citet{Raaijmakers2021} adopting their piecewise-polytropic (PP) and sound-speed (CS) parametrized EoS models, respectively; the dark red line shows the result from \citet{Legred2021} based on a non-parametric, Gaussian-processes-based EoS model; the purple line shows the result from \citet{Pang2021}, who adopt an EoS model informed by chiral effective field theory.}}
  \label{fig:MTOV_prior}
\end{figure}

\begin{figure}
 \centering
 \includegraphics[width=\columnwidth]{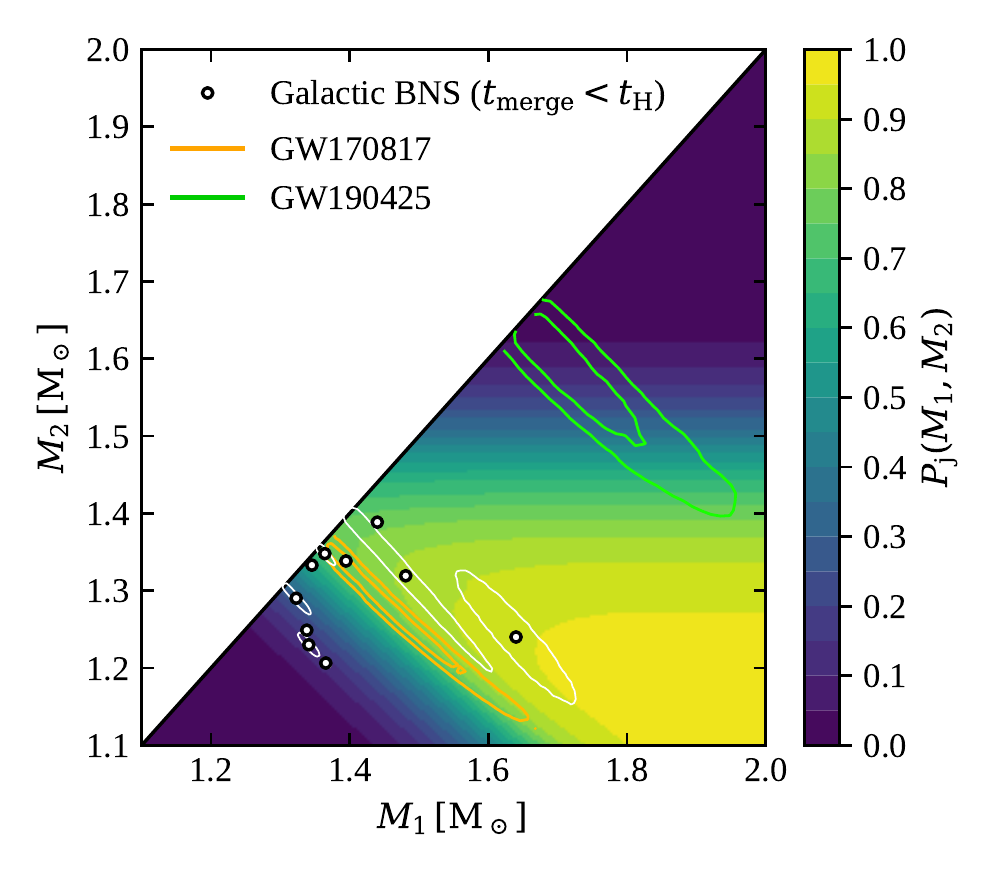}
 \caption{Probability that a BNS would satisfy our jet launching conditions (Eq.~\ref{eq:Pj_m1m2}) as a function of the component masses $M_1$ and $M_2$, given our present knowledge of the EoS (encoded in the priors defined in \S\ref{sec:EoS_priors}). The masses of Galactic BNS systems that merge within a Hubble time and the two GW-detected BNS are shown for comparison, in the same way as in Fig.~\ref{fig:M1_M2_plane}.} 
 \label{fig:Pjet_m1m2}
\end{figure}
 
In our framework, the currently available information on the EoS needs to expressed in the form of priors on $R_{1.4}$ and $M_\mathrm{TOV}$. For both quantities there exists a variety of recent constraints that use either electromagnetic or GW information, or try to combine both. 

The very thorough and complete recent work by \cite{Miller2021} based on X-ray pulsar radius measurements from NICER and XMM-Newton combined with GW constraints found $R_{1.4}=12.45\pm 0.65\,\mathrm{km}$ (one sigma) consistently using three independent frameworks, showing that the result is \resp{robust} against systemtatics. We will therefore use a normal distribution with mean $12.45\,\mathrm{km}$ and standard deviation $0.65\,\mathrm{km}$ as our prior on $R_{1.4}$. 

The available constraints on $M_\mathrm{TOV}$ are more uncertain and model-dependent, as can be partly appreciated by looking at the posteriors on $M_\mathrm{TOV}$ reported in Fig.~\ref{fig:MTOV_prior}, which have been obtained in recent works by three different groups using four different frameworks. On the other hand, a model-independent constraint is set by the mass of the heaviest known pulsar\footnote{\resp{The PSRJ0740+6620 pulsar is inferred to spin at a frequency of roughly 350 Hz \citep{Cromartie2020}, which is up to 25\% of the Keplerian (i.e.\ mass shedding) frequency for viable EoSs \citep[see][]{Haensel2009}. Rotation at such a rate increases the maximum mass that can be supported by the EoS by up to 0.85\% \citep{Breu2016}. We neglect this small correction here, as it is well below our uncertainties.}} PSRJ0740+6620 \citep{Cromartie2020}, whose measurement has been recently refined \citep{Fonseca2021} to $M_\mathrm{PSR}=2.08\pm 0.07\,\mathrm{M_\odot}$ based on high-cadence data from the Green Bank Telescope (GBT) and the Canadian Hydrogen Intensity Mapping Experiment (CHIME). Moreover, \resp{the observation of a jet in GW170817 implies, according to our jet-launching conditions, that the remnant gravitational mass $M_\mathrm{rem,GW17}$ was above $1.2 M_\mathrm{TOV}$. Within our framework, after fixing $R_{1.4}$ we can compute the posterior probability on $M_\mathrm{rem,GW17}$ using the procedure described in \S\ref{sec:remnant_determination} and the publicly available posterior samples on the GW170817 component masses (\citealt{Abbott2019_GW170817_properties}, where we conservatively adopted the samples from the high-spin prior analysis, which yield broader posteriors). Then, the conditions $M_\mathrm{TOV}>M_\mathrm{PSR}$ and $1.2 M_\mathrm{TOV}<M_\mathrm{rem,GW17}$ lead to a prior on $M_\mathrm{TOV}$ conditioned on a particular value of $R_{1.4}$ that can be expressed as} 
 \begin{equation}
 \begin{split}
  & \pi(M_\mathrm{TOV}\,|\,R_{1.4},\xi_\mathrm{d}) \propto\\
  & \Phi_{M_\mathrm{PSR}}(M_\mathrm{TOV})\left[1-\Phi_{M_\mathrm{rem,GW17}}(1.2 M_\mathrm{TOV})\right],
 \end{split}
 \end{equation}
where $\Phi_x(m)$ represents the cumulative posterior distribution of $x$ \resp{evaluated at $m$}. \resp{To account for the uncertainty in the disc mass fitting formula, we introduced in the above formulation an additional nuisance parameter $\xi_\mathrm{d}$, so that $M_\mathrm{disc}(M_2,R_{1.4},\xi_\mathrm{d})=\xi_\mathrm{d}M_\mathrm{disc,KF20}(M_2,R_{1.4})$. The prior on this parameter is defined to reflect the 50\% uncertainy in the disc mass fitting formula as estimated by \citet{Kruger2020}, that is}
\begin{equation}
 \pi(\xi_\mathrm{d})\propto \mathcal{N}_\mathrm{1,0.5}(\xi_\mathrm{d})\Theta(\xi_\mathrm{d}),
\end{equation}
\resp{where}
\begin{equation}
\mathcal{N}_{\mathrm{\mu},\mathrm{\sigma}}(x)\propto \exp\left[-\frac{1}{2}\left(\frac{x - \mu}{\sigma}\right)^2\right].
\end{equation}
\resp{The joint prior on $M_\mathrm{TOV}$, $R_{1.4}$ and $\xi_\mathrm{d}$ is therefore}
\begin{equation}
 \pi(M_\mathrm{TOV},R_{1.4},\xi_\mathrm{d})=\pi(R_{1.4})\pi(\xi_\mathrm{d})\pi(M_\mathrm{TOV}\,|\,R_{1.4},\xi_\mathrm{d}),
 \label{eq:joint_prior}
\end{equation}
where
\begin{equation}
\pi(R_{1.4})=\mathcal{N}_\mathrm{12.45,0.65}\left(\frac{R_\mathrm{1.4}}{\mathrm{km}}\right)
\end{equation}
\resp{as explained above.}   

Figure~\ref{fig:MTOV_prior} shows the \resp{marginalised} prior \resp{on} $M_\mathrm{TOV}$, \resp{that is, Eq.~\ref{eq:joint_prior} marginalised over $R_{1.4}$ and $\xi_\mathrm{d}$} (thick blue solid line), compared to recent multi-messenger constraints on $M_\mathrm{TOV}$ from the literature. \resp{The comparison shows that our prior on $M_\mathrm{TOV}$ is compatible with other recent works in the literature, but with less support at high masses in most cases, as a consequence of the assumption on the GW170817 remnant. We note that similar results based on analogous arguments have been presented by, e.g.,\ \citet{Rezzolla2018}, \citet{Shibata2019} and \citet{Annala2022}.}

Using the \resp{above-defined joint} prior, we can compute the probability that a $M_1$, $M_2$ binary satisfies our jet launching conditions as
 \begin{equation}
  P_{\mathrm{j}}(M_1,M_2) = \iiint \Theta_\mathrm{j}\pi(M_\mathrm{TOV},R_{1.4},\xi_\mathrm{d})\,\mathrm{d}M_\mathrm{TOV}\,\mathrm{d}R_{1.4}\,\mathrm{d}\xi_\mathrm{d}
  \label{eq:Pj_m1m2}
 \end{equation}
where $\Theta_\mathrm{j}(M_1,M_2,\vec\theta_\mathrm{EoS})=1$ if our jet launching conditions are met, and $\Theta_\mathrm{j}=0$ otherwise. More explicitly, 
\begin{equation}
 \Theta_\mathrm{j}=\Theta(M_\mathrm{rem}-1.2M_\mathrm{TOV})\Theta(M_\mathrm{disc}-M_\mathrm{disc,min}),
 \label{eq:Theta_j}
\end{equation}
\resp{where} we stress that $M_\mathrm{rem}$ and $M_\mathrm{disc}$ depend on the merging binary component masses, the EoS \resp{and the nuisance parameter $\xi_\mathrm{d}$}. The result of Eq.~\ref{eq:Pj_m1m2} for our choice of EoS priors is shown in Figure~\ref{fig:Pjet_m1m2}. The lighter the colour, the more likely our jet launching conditions are met, given our present knowledge about the EoS\footnote{\resptwo{We note that the value of $P_\mathrm{j}$ averaged over the GW170817 joint component mass posterior \citep[adopting the low-spin-prior samples from][]{Abbott2019_GW170817_properties} is 78\%. This is not 100\% because the information that GW170817 successfully launched a jet enters only through the prior on $M_\mathrm{TOV}$ and it cannot be applied self-consistently to the GW170817 masses within the present method. An alternative approach that solves this issue is outlined in the discussion section.}}. For systems located in the dark lower-left corner of the plot, we can confidently predict that their remnant will not collapse to a black hole on a short timescale, preventing the Blandford-Znajek mechanism from operating. \resp{Jet-launching by the same mechanism is also confidently excluded for systems in the upper-right dark corner, as these produce negligible accretion discs.} The $f_\mathrm{j,tot}$ (or $f_\mathrm{j,GW}$) lower limits from section \ref{sec:fjet_estimation} therefore lead qualitatively to the conclusion that \textit{we can exclude that the large majority \resp{(indicatively more than 90\%)} of BNS systems are located in the darkest regions of this plot.} The absence of observed BNS mergers or Galactic double neutron stars in these regions is therefore not due to selection effects, and models that predict a large fraction of low-mass systems, such as many population synthesis models (e.g.~\citealt{Fryer2012,Dominik2012,VignaGomez2018,Mapelli2018}), are therefore in clear tension with these results.


\section{Inference on the BNS mass distribution and the EoS based on the jet incidence}

\subsection{Derivation}
\label{sec:derivation}

Our jet launching conditions depend on the component masses $M_1$, $M_2$ and on the EoS through $M_\mathrm{TOV}$ and the dependence of the disc mass on the \resp{compactness of the secondary} neutron star. \resp{For a fixed} mass distribution $\mathrm{d}^2P/\mathrm{d}M_1\mathrm{d}M_2$ of BNS mergers and EoS, \resp{one can compute} $f_\mathrm{j,tot}$ directly from
\begin{equation}
 \tilde f_\mathrm{j,tot} = \iint\frac{\mathrm{d}^2P}{\mathrm{d}M_1\mathrm{d}M_2} \Theta_\mathrm{j}\,\mathrm{d}M_1\mathrm{d}M_2,
\end{equation}
where we use the tilde ($\,\tilde ~\,$) to distinguish this functional (which depends on the mass distribution and on the EoS) from the $f_\mathrm{j,tot}$ \resp{observable}. Similarly, \resp{one} could compute the jet \resp{incidence} in GW-detected binaries as
\begin{equation}
 \tilde f_\mathrm{j,GW} = \frac{\iint V_\mathrm{eff}\frac{\mathrm{d}^2P}{\mathrm{d}M_1\mathrm{d}M_2}\Theta_\mathrm{j}\,\mathrm{d}M_1\mathrm{d}M_2}{\iint V_\mathrm{eff}\frac{\mathrm{d}^2P}{\mathrm{d}M_1\mathrm{d}M_2}\mathrm{d}M_1\mathrm{d}M_2},
 \label{eq:ftildejGW}
\end{equation}
where $V_\mathrm{eff}$ represents the GW detector sensitive volume for a binary with masses $M_1$ and $M_2$, averaged over all extrinsic parameters. For the present detector network, whose BNS range is bound to $z\ll 1$ and whose SNR is dominated by the inspiral phase, a good approximation is $V_\mathrm{eff}\propto M_c^{5/2}$, where $M_c=(M_1M_2)^{3/5}/(M_1+M_2)^{1/5}$ is the chirp mass (e.g.~\citealt{Mandel2019}). In what follows, we will use $f_\mathrm{j}$ to indicate either $f_\mathrm{j,tot}$ or $f_\mathrm{j,GW}$ (depending on which one of the constraints shown in Fig.~\ref{fig:cumulative_posterior} is used), and $\tilde f_\mathrm{j}$ to indicate $\tilde f_\mathrm{j,tot}$ or $\tilde f_\mathrm{j,GW}$ accordingly; we will also use $\vec d$ to indicate the data that is used to constrain $f_\mathrm{j}$, that is the information on the jet \resp{incidence} in GW-detected BNS binaries \resp{(\S\ref{sec:fjet_GW}, i.e.\ the fact that GW170817 launched a jet, \resp{which we denote hereon by $\vec d_\mathrm{GW}$})} or that on the SGRB and BNS local rate densities \resp{(\S\ref{sec:fj_from_R0}, i.e.\ $\vec d_\mathrm{tot}$)}.  Expressing the mass distribution in a parametric form with a set of parameters $\vec \theta_\mathrm{m}$, and again the EoS with a set of parameters $\vec \theta_\mathrm{EoS}$ \resp{(we include in this set, for ease of presentation, also the nuisance parameter $\xi_\mathrm{d}$)}, we can formally use Bayes' theorem to infer posteriors on these parameters from $f_\mathrm{j}$, obtaining
\begin{equation}
 P(\vec \theta_\mathrm{m},\vec \theta_\mathrm{EoS}|\,f_\mathrm{j})
 \propto P(f_\mathrm{j}\,|\,\vec \theta_\mathrm{m},\vec \theta_\mathrm{EoS})\pi(\vec \theta_\mathrm{m},\vec \theta_\mathrm{EoS}),
\end{equation}
where $\pi(\vec\theta_\mathrm{m},\vec \theta_\mathrm{EoS})$ is the prior\footnote{The priors on $\vec \theta_\mathrm{m}$ and $\vec \theta_\mathrm{EoS}$ are not necessarily independent from each other, as the mass distribution parameters can for example include $M_\mathrm{TOV}$.} on $\vec\theta_\mathrm{m}$ and $\vec\theta_\mathrm{EoS}$, and the likelihood term is simply 
\begin{equation}
 P(f_\mathrm{j}\,|\,\vec \theta_\mathrm{m},\vec \theta_\mathrm{EoS}) = \delta(f_\mathrm{j}-\tilde f_\mathrm{j}(\vec \theta_\mathrm{m},\vec \theta_\mathrm{EoS})).
 \label{eq:fjGWdelta}
\end{equation}
Since the latter is not known exactly, but it is in turn inferred from $\vec d$, we have
\begin{equation}\label{eq:mass_and_eos_inference}
 P(\vec \theta_\mathrm{m},\vec \theta_\mathrm{EoS}\,|\,\vec d) = \int_0^{1} P(\vec \theta_\mathrm{m},\vec \theta_\mathrm{EoS}\,|\,f)P(f_\mathrm{j}\,|\,\vec d)\,\mathrm{d}f_\mathrm{j},
\end{equation}
which, using Eq.~\ref{eq:fjGWdelta}, simplifies to
\begin{equation}
 P(\vec \theta_\mathrm{m},\vec \theta_\mathrm{EoS}\,|\,\vec d) \propto P\left(f_\mathrm{j}=\tilde f_\mathrm{j}(\vec \theta_\mathrm{m},\vec \theta_\mathrm{EoS})\,|\,\vec d\right)\pi(\vec\theta_\mathrm{m},\vec\theta_\mathrm{EoS}).
 \label{eq:P(m,EoS)_general}
\end{equation}
This equation expresses the joint constraint on the EoS and mass distribution parameter space that can be set by our arguments. For current constraints, using equations \ref{eq:fjGWposterior} and \ref{eq:P(fj from R0)} and \resp{conservatively} adopting a uniform prior on $f_\mathrm{j}$, this can be finally written as
\begin{equation}   
 P(\vec \theta_\mathrm{m},\vec \theta_\mathrm{EoS}\,|\,\vec d) \propto\left [ \tilde f_\mathrm{j}(\vec \theta_\mathrm{m},\vec \theta_\mathrm{EoS})\right ]^\gamma \pi(\vec\theta_\mathrm{m},\vec\theta_\mathrm{EoS}).
 \label{eq:P(m,EoS)} 
\end{equation}
where $\gamma=1$ if $f_\mathrm{j}=f_\mathrm{j,GW}$ (see \S\ref{sec:f_jet_min}) or $\gamma=0.42$ if $f_\mathrm{j}=f_\mathrm{j,tot}$ (see \S\ref{sec:fj_from_R0}). The somewhat shallower constraint on $f_\mathrm{j,tot}$ with respect to $f_\mathrm{j,GW}$ is partly compensated by the $V_\mathrm{eff}$ term in Eq.~\ref{eq:ftildejGW} but, as we show in the following section, the $f_\mathrm{j,GW}$ is still the most constraining one between the two, given the currently available data.

\subsection{Mass distribution constraints}
\label{sec:mass_distrib_constraints}
\begin{figure} 
 \centering
 \includegraphics[width=\columnwidth]{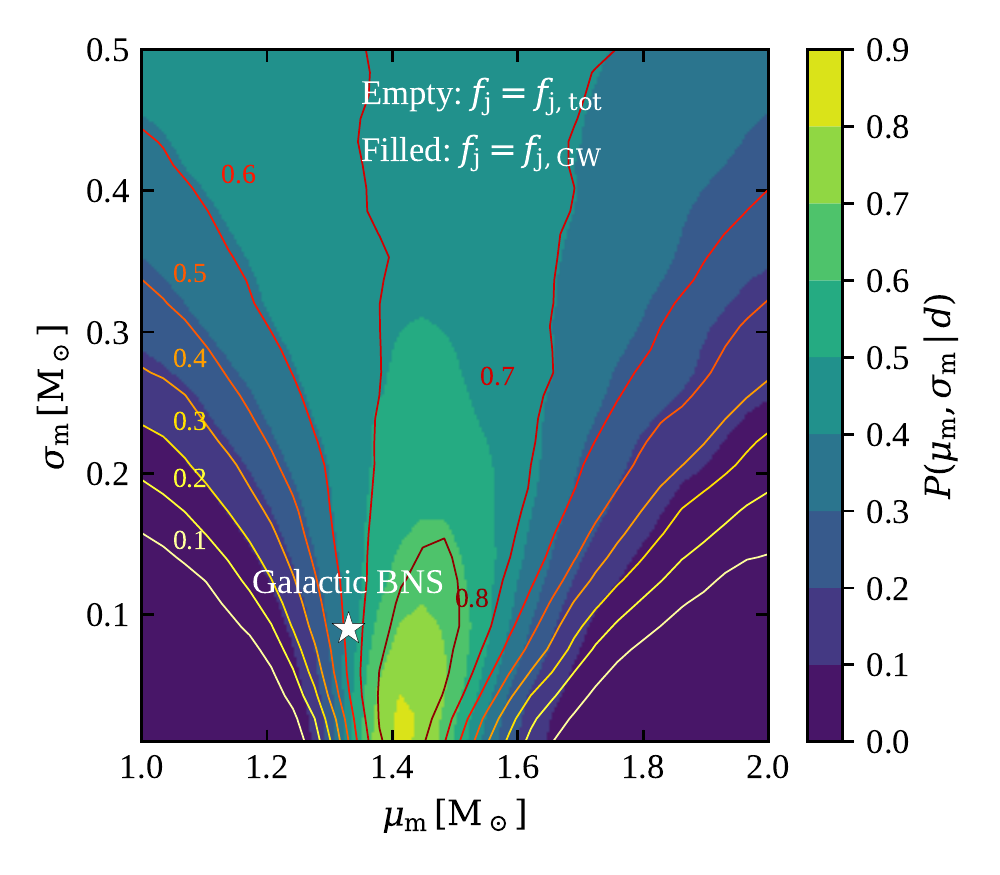}
 \caption{Constraints on the parameters of a Gaussian BNS mass distribution model. Filled contours show the posterior probability density on the $\mu_\mathrm{m}$ (mean) and $\sigma_\mathrm{m}$ (standard deviation) parameters of a Gaussian mass distribution model (assuming random pairing) obtained using $f_\mathrm{j}=f_\mathrm{j,GW}$, i.e.~the constraint based on the association of GW170817 and GRB170817A (\S\ref{sec:fjet_estimation}). Empty contours show the corresponding result for $f_\mathrm{j}=f_\mathrm{j,tot}$, i.e.~the constraint based on the relative SGRB and BNS local rate densities (\S\ref{sec:fj_from_R0}). The white star marks parameter values $(\mu_\mathrm{m},\sigma_\mathrm{m})=(1.33,0.09)\,\mathrm{M_\odot}$, representative of the main peak of the observed Galactic BNS mass distribution \citep{Ozel2016}.}  
 \label{fig:Pm_mu_sigma}
\end{figure}

The formalism derived in the previous section can be used to investigate the parameter space of a mass distribution model. Given our a priori knowledge of the EoS, encoded in the prior, we can marginalise Eq.~\ref{eq:P(m,EoS)} over $\vec\theta_\mathrm{EoS}$ to formally obtain the posterior on the mass distribution parameters, namely
\begin{equation}
 P(\vec\theta_\mathrm{m}\,|\,\vec d)\propto\int\tilde f_\mathrm{j}^\gamma \,\pi(\vec\theta_\mathrm{EoS},\vec\theta_\mathrm{m})\,\mathrm{d}\vec\theta_\mathrm{EoS}.
 \label{eq:P(m)}
\end{equation}
Taking a uniform prior on the mass distribution parameters, $\pi(\vec\theta_\mathrm{m})\propto 1$, this essentially reduces to averaging $\tilde f_\mathrm{j}^\gamma$ over the EoS uncertainty for a fixed choice of $\vec\theta_\mathrm{m}$.
Figure~\ref{fig:Pm_mu_sigma} shows the latter quantity for a simple mass distribution model where both binary component masses are sampled from the same Gaussian distribution with mean $\mu_\mathrm{m}$ and $\sigma_\mathrm{m}$. \resp{We consider wide ranges for these parameters, that is $1\leq \mu_\mathrm{m}/\mathrm{M_\odot}\leq 2$ (encompassing the range of observed neutron star masses) and $0.01\leq\sigma_\mathrm{m}/\mathrm{M_\odot}\leq 0.5$ (from very narrow to very wide Gaussians over the considered mass range).} The marginalisation over the EoS parameters is performed using our simplified two-parameter EoS dependence (\S\ref{sec:approx_bzconditions}), adopting the priors introduced in \S\ref{sec:EoS_priors}, limiting the NS masses to the range $M_{1,2}\in[1\,\mathrm{M_\odot},M_\mathrm{TOV}]$. Filled contours show the result obtained using $f_\mathrm{j}=f_\mathrm{j,GW}$, while empty contours show the corresponding result when adopting the $f_\mathrm{j}=f_\mathrm{j,tot}$ alternative constraint. 
This shows that a narrow ($\sigma_\mathrm{m}\lesssim 0.2\,\mathrm{M_\odot}$) Gaussian distribution with $\mu_\mathrm{m}\lesssim 1.3\,\mathrm{M_\odot}$ or $\mu_\mathrm{m}\gtrsim 1.6\,\mathrm{M_\odot}$ is strongly disfavoured due to the very low implied jet \resp{incidence}, \resp{given the available information on} the EoS \resp{and the jet incidence}, and consistently when using either of the constraints we considered (\resp{that is}, $\vec d_\mathrm{GW}$ or $\vec d_\mathrm{tot}$). On the other hand, parameters representative of the main peak of the observed Galactic BNS population (white star in Fig.~\ref{fig:Pm_mu_sigma}) are \resp{consistent with the constraints}. 

Given the fact that the analysis of GW-detected BNS \citep{LVC2021_GWTC3pop,Landry2021}, radio-detected Galactic BNS \citep{Farrow2019} and both populations together \citep{Galaudage2021} currently disfavour a narrow -peaked mass distribution, we also tested a different parametric form, that is, a power law mass distribution model where both components are sampled from a power law $P(m\,|\,\alpha,M_\mathrm{min},M_\mathrm{TOV})\propto m^\alpha$ with $M_\mathrm{min}\leq m \leq M_\mathrm{TOV}$, where $m$ is the mass of either component. \resp{We consider a wide range of slopes $-20\leq \alpha \leq 10$, and we take $1\leq M_\mathrm{min}/\mathrm{M_\odot}\leq 1.25$, with the upper bound being limited by the lightest observed masses Galactic neutron stars \citep{Farrow2019}}.
\begin{figure}
 \centering \includegraphics[width=\columnwidth]{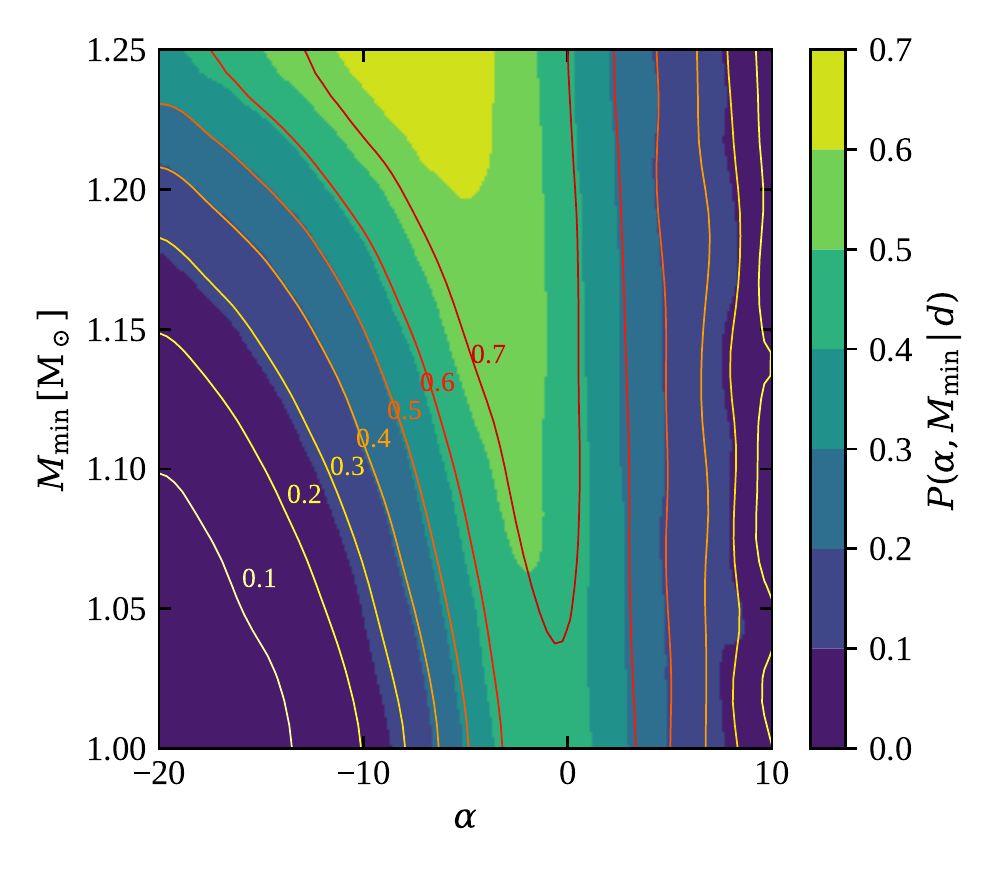}
 \caption{Constraints on the parameters of a power law BNS mass distribution model. Colours and lines have the same meaning as in Fig.~\ref{fig:Pm_mu_sigma}.}
 \label{fig:Pm_a_Mmin}
\end{figure}
Figure~\ref{fig:Pm_a_Mmin} shows the resulting constraint on the $(\alpha,M_\mathrm{min})$ plane, with the same colour coding as for the Gaussian mass distribution. 

It is instructive to project both results on the actual mass distribution space. We stress that we assume the component masses to be sampled from the same distribution (and randomly paired) which we refer to as $P_m(m\,|\,\theta_\mathrm{m})$. The implied primary mass distribution is therefore
\begin{equation}
\begin{split}
& P(M_1\,|\,\theta_\mathrm{m})= \\
& 2\int P_m(M_1\,|\,\theta_\mathrm{m})P_m(M_2\,|\,\theta_\mathrm{m})\Theta(M_1-M_2)\,\mathrm{d}M_2, 
\end{split}
\end{equation}
and similarly for the secondary mass distribution.  We keep the same uniform priors as in Figures~\ref{fig:Pm_mu_sigma} and \ref{fig:Pm_a_Mmin},  and we use the constraint on $f_\mathrm{j}=f_\mathrm{j,GW}$.
\begin{figure*}[t]
 \centering \includegraphics[width=\textwidth]{figures/dP_dm1_projected.pdf}
 \caption{Mass distribution constraint. In each panel, filled  contours show the 90\% and 50\% credible regions of the probability distribution $P(m\,|\,\theta_\mathrm{m})$ from which both $M_1$ and $M_2$ are sampled, while the thick solid line shows the median distribution at each $m$. The $\vec d_\mathrm{j,GW}$ constraint is used. Grey dashed lines show the edges of the corresponding 90\% and 50\% credible regions when using only the priors. The left-hand panel shows the result for the power law mass distribution model, the central panel shows that for the Gaussian model, and the right-hand panel compares the two results with a log-scale vertical axis, to emphasise the lower 90\% contour. }
 \label{fig:dP_dm1_projected}
\end{figure*}
Figure~\ref{fig:dP_dm1_projected} shows the resulting mass distribution constraints. \resp{The credible bands in the figure are obtained as follows: we sampled the $\theta_\mathrm{m}$ posterior (computed according to Eq.~\ref{eq:P(m)}) and constructed the $P(m\,|\,\theta_\mathrm{m})$ curves corresponding to the samples. For each fixed value of $m$, the 50\% (resp.\ 90\%) credible band is comprised between the 25$^\mathrm{th}$ and the 75$^\mathrm{th}$ (resp.\ 5$^\mathrm{th}$ and the 95$^\mathrm{th}$) percentiles of the values of these curves at $m$. Similarly, the thick solid lines represent the medians of these values}. The cut-offs below $1\,\mathrm{M_\odot}$ and above $2.2\,\mathrm{M_\odot}$ are determined by our priors; in between this range, the result shows some improvement with respect to the priors and generally disfavours narrow mass distributions. The clearest feature, common to the two parametrisations (as is particularly visible in the comparison panel), is the requirement of a non-negligible probability for masses in the $1.3-1.6\,\mathrm{M_\odot}$ range, which are particularly well-positioned in terms of satisfying our jet-launching conditions, as shown in Figure~\ref{fig:Pjet_m1m2}. This is in good agreement with the position and width of the main mass peak as found by \citet{Galaudage2021}. In Appendix~\ref{sec:GWTC3_comparison} we show a comparison of these results to those from the recently circulated preprint on the LIGO/Virgo GWTC-3 population study by \citet{LVC2021_GWTC3pop}.
 
\subsection{EoS constraints}
\label{sec:EoS_constraints}
\begin{figure} 
 \centering
 \includegraphics[width=\columnwidth]{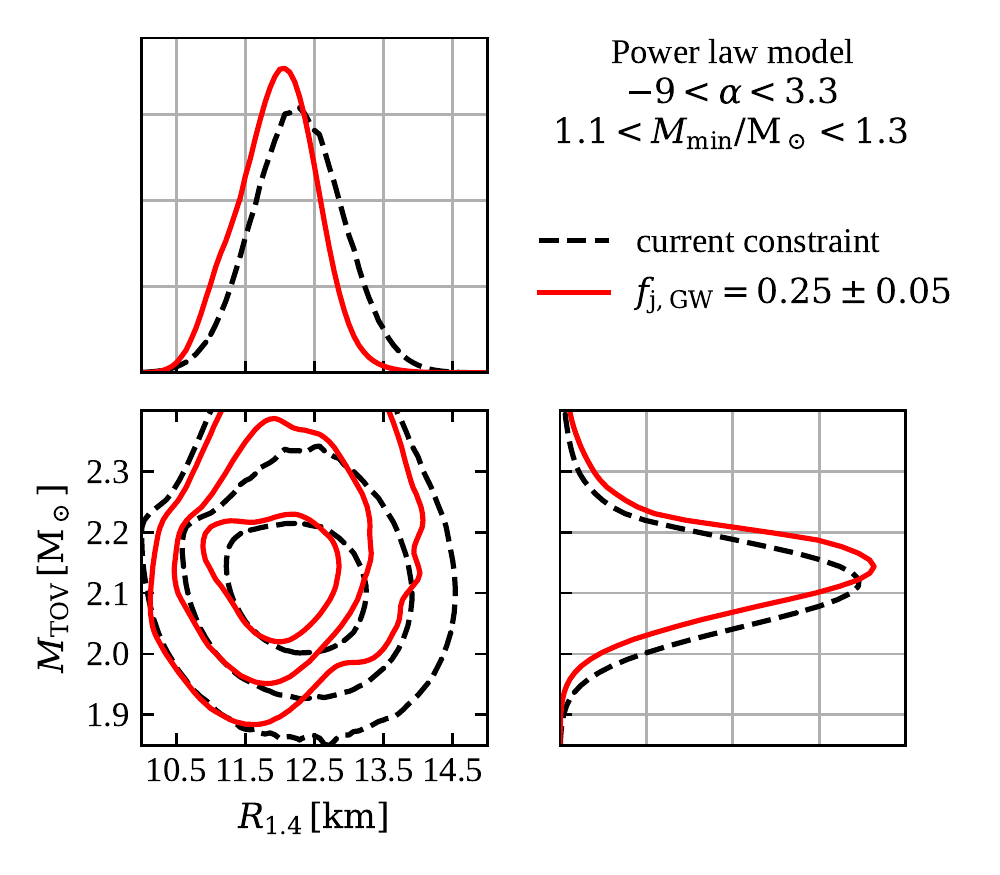}
 \caption{Illustrative constraints on the EoS parameters assuming \resp{the power law mass distribution model with uniform priors on $1.1\leq M_\mathrm{min}/\mathrm{M_\odot}\leq 1.3$ and $-9\leq \alpha\leq 3.3$, matching the 90\% credible ranges of the GWTC-3 NS mass distribution constraint \citep{LVC2021_GWTC3pop}, and using a hypothetical future constraint on the jet incidence $P(f_\mathrm{j,GW}\,|\,\vec d_\mathrm{GW})\propto \mathcal{N}_{0.25,0.05}(f_\mathrm{j,GW})$}.}  
 \label{fig:PEoS_fjknown025} 
\end{figure}  
The posterior in Eq.~\ref{eq:P(m,EoS)} can also be marginalised over the mass distribution parameters, to obtain constraints on the EoS given the knowledge of the mass distribution. Unfortunately, the current uncertainties on the jet \resp{incidence} and mass distribution are too large, and this simply returns the EoS priors (which are already quite informative).

\begin{figure}
 \centering
 \includegraphics[width=\columnwidth]{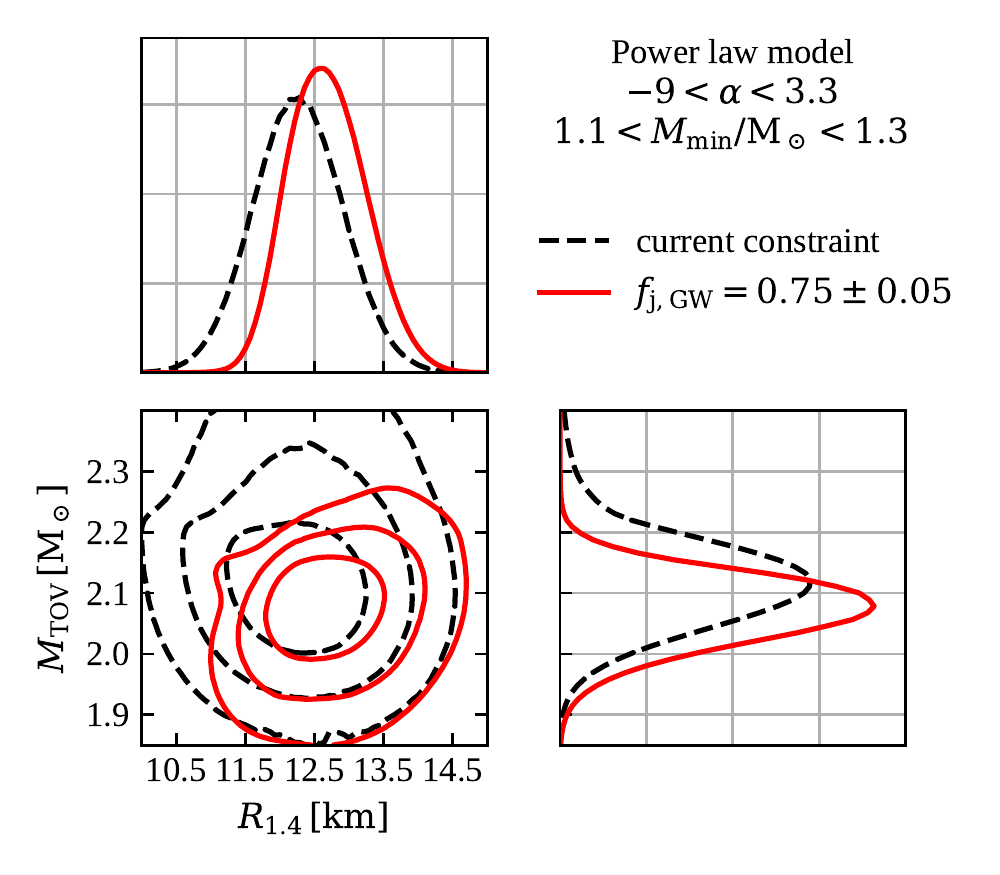}
 \caption{\resp{Same as Figure~\ref{fig:PEoS_fjknown025}, but assuming a jet incidence constraint with a different mean, that is $P(f_\mathrm{j,GW}\,|\,\vec d_\mathrm{tot})\propto \mathcal{N}_{0.75,0.05}(f_\mathrm{j,GW})$}.}  
 \label{fig:PEoS_fjknown075}
\end{figure}

To \resp{give a basic illustration of the relevance of this kind of constraint in the future,} we consider \resp{the following example}. We adopt the power law mass distribution model \resp{described above}, with a uniform prior on $M_\mathrm{min}$ between $1.0\,\mathrm{M_\odot}$ and $1.3\,\mathrm{M_\odot}$, and a uniform prior on $\alpha$ between $-9$ and $3.3$, \resp{matching the 90\% credible ranges from the GWTC-3 population study \citep{LVC2021_GWTC3pop}}. \resp{This therefore roughly represents the current uncertainty on the NS mass distribution in merging stellar-mass compact object binaries. As a way of representing a hypothetical future in which the jet incidence is known with relatively small uncertainty, we} assume the posterior probability on the jet \resp{incidence in GW-detectable binaries} $f_\mathrm{j,GW}$ to be represented by a Gaussian with standard deviation $\sigma_{f}=0.05$ \resptwo{(which represents roughly a 6-fold improvement over the current uncertainty, see Fig.~\ref{fig:cumulative_posterior})} \resp{and two possible mean values, $\mu_f=0.25$ and $\mu_f=0.75$}. Eq.~\ref{eq:P(m,EoS)_general} then becomes
\begin{equation}
\begin{split}
 & P(\alpha,M_\mathrm{min},M_\mathrm{TOV},R_{1.4},\xi_\mathrm{d}\,|\,\vec d_\mathrm{GW})\propto \\
 & \propto\mathcal{N}_{\mathrm{\mu_f},\mathrm{\sigma_f}}(\tilde f_\mathrm{j,GW})\pi(M_\mathrm{TOV},R_{1.4},\xi_\mathrm{d})\pi(\alpha)\pi(M_\mathrm{min}).
\end{split} 
\end{equation}
The marginalised distributions $P(M_\mathrm{TOV},R_{1.4}\,|\,\vec d_\mathrm{GW})$ \resp{for the two choices of $\mu_\mathrm{f}$} are shown in Figures~\ref{fig:PEoS_fjknown025} \resp{and \ref{fig:PEoS_fjknown075}}.
The results 
show that \resp{the posterior probability on the EoS parameters is sensitive to the value of $f_\mathrm{j,GW}$} even when the BNS mass distribution is \resp{relatively} poorly known.
We conclude that in the near future, when the BNS mass distribution (plus the local rate and possibly the jet \resp{incidence}) will be better constrained thanks to more GW observations, this methodology will be able to place interesting constraints on the EoS parameters.
\resp{This is in agreement with \citealt{Fryer2015}, who first discussed the possiblity to constrain $M_\mathrm{TOV}$ from the jet incidence in GW-detectable BNS mergers.}

\section{Discussion and conclusions}
\label{sec:discussion}


The successful jet in GW170817 not only provided \resp{strong evidence in support} of the long-held binary neutron star progenitor scenario for short gamma-ray bursts, but also clearly indicated that \resp{the launch of a relativistic jet} is  most likely a common outcome of this kind of mergers. In this work, we have shown how the fraction of BNS mergers that launch an ultra-relativistic jet can be used to place a joint constraint on the binary neutron star mass distribution and on the neutron star matter equation of state, under the assumption that the jet-launching mechanism requires the presence of an accretion disc with a non-negligible mass and the collapse of the merger remnant to a black hole within a fraction of a second (i.e.~a supra-massive or stable neutron star remnant are assumed not to yield SGRBs). 

Based on the presence of a jet in GW170817 and adopting a simple binomial likelihood and a uniform prior, we derived a lower limit $f_\mathrm{j,GW}>32\%$ (90\% credible level) on the \resp{incidence} of jets in GW-detected binary neutron stars. By comparing the lower limit on the local short gamma-ray burst rate density that can be placed by the detection of GRB170817A by \textit{Fermi}/GBM and the local binary neutron star merger rate density estimated by \citet{Grunthal2021} based on Galactic double neutron stars, we constrained the jet \resp{incidence} in the whole population to be $f_\mathrm{j,tot}>20\%$.  This constraint is weaker than the finding of $f_\mathrm{j,tot}> 40\%$ (90\% credible level) by \citet{Sarin2022}, a pre-print \resp{-- now published --}  that appeared while this work was nearing submission; however, the quantitative discrepancy is likely due to the reliance by \citet{Sarin2022} on the inferred local SGRB rate density estimated by \citet{Coward2012} and \citet{Wanderman2015} -- which most likely suffer from systematics due to the uncertain luminosity function (especially at the low end) and redshift distribution of events (see \S\ref{sec:fj_from_R0}) -- and their assumption that promptly collapsing BNS merger products without a signifiant accretion disc could still launch a jet and power a SGRB.

Comparing these results with the predicted jet \resp{incidence} in a \citet{Blandford1977} jet-launching scenario, assuming equation of state priors informed by the latest multi-messenger constraints, we can exclude mass distributions with an overwhelming fraction of low-mass ($\lesssim 1.3\,\mathrm{M_\odot}$) neutron stars (as predicted by many population synthesis models, especially those which adopt the \citealt{Fryer2012} `rapid' supernova prescription).  This finding qualitatively agrees with \citealt{Sarin2022} (their Figure 1). Similarly, a mass distribution dominated by high-mass ($\gtrsim 1.6\,\mathrm{M_\odot}$) components is strongly disfavoured. We caution that this conclusion depends critically on the assumed jet-launching conditions: if an alternative magnetar central engine scenario (though theoretically less favoured) was adopted, these conclusion would be significantly altered.

The method can be also used in principle to place constraints on the equation of state, but the current data are insufficient to make this methodology competitive. Nevertheless, \resp{in \S\ref{sec:EoS_constraints}} we \resp{provided a first illustration of how} this will improve with future tighter constraints on the jet \resp{incidence}.

\resp{Given the relatively weak constraints that we obtain with the currently available data, this work must be regarded primarily as a proof of concept.} In the near future, several new binary neutron stars are expected \citep{LVC2020prospects} to be detected through gravitational waves in the upcoming O4 and O5 observing runs of the GW  detector network (which now includes KAGRA, \citealt{Somiya2012}). In addition to that, new binary pulsars are being discovered in radio surveys \citep{Han2021,Pan2021,Good2021,Agazie2021} at an increasing rate thanks to the steady improvements in both technology and data analysis techniques.  Last, but not least, our understanding of selection effects that shape the properties of the observed radio pulsar population is advancing \citep{Chattopadhyay2020}. \resp{All these advances are going to positively impact the constraints on $f_\mathrm{j,GW}$ and $f_\mathrm{j,tot}$ (Fig.~\ref{fig:cumulative_posterior}) and the $\theta_\mathrm{m}$ and $\theta_\mathrm{EoS}$ constraints that can be derived through the presented methodology.}

New gravitational-wave detections of binary neutron star mergers will also open the possibility of directly using the information on the presence of a jet in each event (see Appendix~\ref{sec:fjGW_multiple} for an extension to uncertain jet associations) in population studies. In a modelling framework such as that described by \cite{Mandel2019}, \resp{the jet-launching condition (represented by the $\Theta_\mathrm{j}$ term, Eq.~\ref{eq:Theta_j}) could be used, in conjunction with the available information on the presence or absence of a jet in each GW-detected BNS merger, to constrain the parameters of single events in the population and, as a result, of global population parameters (i.e. hyperparameters) such as $\theta_\mathrm{m}$ and $\theta_\mathrm{EoS}$.} This can complement methods that exploit information on kilonova ejecta masses \citep{Li1998,Metzger2019} and SGRB afterglow observations \resp{to augment the information that can be obtained from the} gravitational-wave data \citep[e.g.][]{Hotokezaka2019,Radice2019,Lazzati2019,Barbieri2019,Coughlin2019,Dietrich2020,Breschi2021,Raaijmakers2021_GW190425}. \resp{In this context, multi-messenger observations of BNS mergers close to the boundaries of the currently viable parameter space for jet launching would be particularly informative: as an example, the obsevation of a clear jet signature (or a tight constraint on its presence) in a system whose masses lie in the region where jet-launching is most uncertain (regions where $P_\mathrm{j}\sim 0.5$ in Fig.~\ref{fig:Pjet_m1m2}) would have the highest impact on the results that can be obtained with the presented method. Conversely, observations that clearly contradict the jet-launching conditions adopted in this work (such as a jet-less system with masses in the $P_\mathrm{j}\sim 1$ region in Fig.~\ref{fig:Pjet_m1m2}, or a jet-launching system with masses in the dark blue regions of the same figure) would point to the need for a revision in the assumed jet-launching conditions}.  

\begin{acknowledgements} The authors thank the anonymous referee for a very detailed and insightful report that helped in improving the presentation of the results. The authors also thank Sylvia Biscoveanu and Samuele Ronchini for insightful comments and suggestions. OS thanks Giancarlo Ghirlanda, Monica Colpi, Michele Ronchi, Annalisa Celotti and Floor Broekgaarden for helpful comments and discussions. OS acknowledges financial support from the INAF-Prin 2017 (1.05.01.88.06) and the Italian Ministry for University and Research grants 1.05.06.13 and 20179ZF5KS. IM acknowledges support from the Australian Research Council Centre of Excellence for Gravitational  Wave  Discovery  (OzGrav), through project number CE17010004.  IM is a recipient of the Australian Research Council Future Fellowship FT190100574.  
\end{acknowledgements}

\bibliographystyle{aa}
\footnotesize
\bibliography{references}
\normalsize

\begin{appendix}
\section{Inference on $f_\mathrm{j,GW}$ from multiple observations with uncertain jets}\label{sec:fjGW_multiple}

\begin{figure}
 \centering\includegraphics[width=\columnwidth]{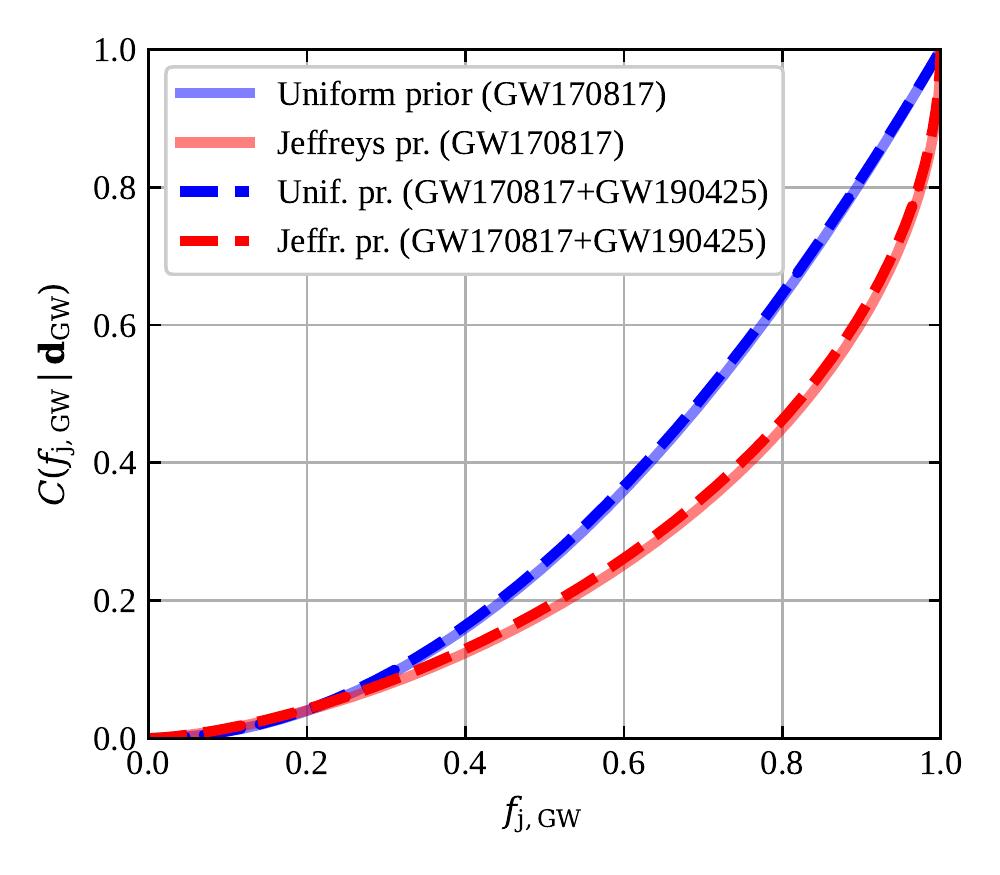}
 \caption{\small Cumulative posterior probability of the BNS jet \resp{incidence}, showing the effect of adding GW190425 to the sample. The solid lines are the same as in Fig.~\ref{fig:cumulative_posterior} and account for GW170817 only, while the dashed ones account for GW190425 as well, as described in the text.}
 \label{fig:fjGW_multiple} 
\end{figure}

In this appendix, we describe how the posterior probability on $f_\mathrm{j,GW}$ can be derived in presence of multiple observations, including cases in which the presence of a jet is uncertain. Let us assume that $n$ events have been observed, and that we are able to estimate, for each $i$-th event, the probability $P_{\mathrm{miss},i}$ that a jet would have been missed (i.e.~no conclusive statement can be made) if present: this will be typically a model-dependent estimate based on the (viewing-angle-dependent) expected jet emission properties, combined with information on the fraction of the BNS merger GW localisation volume that has been surveyed, and the depth of the available observations\footnote{We note that the available observations themselves could have been performed in an attempt to collect conclusive evidence in favour of a jet, in case preliminary observations provided promising indications of its presence. In that sense, $P_\mathrm{miss}$ can evolve as the data are collected and, in favourable cases such as GW170817, it can get close to zero (i.e.~any jet in association to GW170817, if present, would not have been missed, also factoring in the indications towards a relatively small viewing angle that stem from kilonova observations, see e.g.~\citealt{Breschi2021}). This latter statement clearly depends also on the assumed distribution of possible jet luminosities and properties in the probed bands. }. Without using other information (such as that on the single-event masses), the intrinsic probability of the presence of a jet in each GW-detected event is just $f_\mathrm{j,GW}$. The likelihood of observing a jet in association to the $i$-th event is therefore $P_{\mathrm{j},i}=f_\mathrm{j,GW}(1-P_{\mathrm{miss},i})$, while that of observing no jet is $P_{\neg\mathrm{j},i}=(1-f_\mathrm{j,GW}) + f_\mathrm{j,GW}P_{\mathrm{miss},i}=1-P_{\mathrm{j},i}$, so that we can write 
\begin{equation}
 P(J_i\,|\,f_\mathrm{j,GW},P_{\mathrm{miss},i}) = J_i P_{\mathrm{j},i} + (1-J_i) (1-P_{\mathrm{j},i}),
\end{equation}
where $J_i=1$ if conclusive evidence in favour of a jet was found in association to the $i$-th event, and $J_i=0$ otherwise.
The $f_\mathrm{j,GW}$ posterior, up to a normalization factor, is then
\begin{equation}
 P(f_\mathrm{j,GW}\,|\,\vec J,\vec P_\mathrm{miss}) \propto \pi(f_\mathrm{j,GW})\prod_{i=1}^{n} P(J_i\,|\,f_\mathrm{j,GW},P_{\mathrm{miss},i}),
 \label{eq:P(fjGW | J,Pmiss)} 
\end{equation}
where $\vec J=\lbrace J_1,J_2,...,J_n\rbrace$ and $\vec P_\mathrm{miss}=\lbrace P_{\mathrm{miss},1},P_{\mathrm{miss},2},...,P_{\mathrm{miss},n}\rbrace$.
Let us evaluate Eq.~\ref{eq:P(fjGW | J,Pmiss)} in a concrete case, that of GW170817 and GW190425. In this setting, we have $n=2$ and $\vec J=(1,0)$. Given the short distance to GW170817, its precise localization, and the availability of an extensive multi-wavelength dataset, including VLBI observations, we can set $P_{\mathrm{miss},1}\sim 0$. In the case of GW190425, on the other hand, \resp{the poor GW sky localisation and larger distance hampered the detection and identification of a putative kilonova, which would have led to a precise localisation enabling deep multi-wavelength monitoring. For that reason,} the available observations can only tentatively exclude an on-axis jet \citep{Hosseinzadeh2019}. In particular, gamma-ray upper limits set by \textit{Fermi}/GBM, \textit{INTEGRAL} and Konus-\textit{Wind} (the latter covering the entire GW localization region of GW190425) can be interpreted as the indication that the viewing angle of a putative jet, if present, must have been sufficiently large. Assuming that any jet would have been missed beyond a limiting viewing angle $\theta_\mathrm{v,lim}$, and calling $P(\theta_\mathrm{v})$ the viewing angle probability distribution of GW190425, a relativistic jet, if present, would have been missed with a probability
\begin{equation}
 P_{\mathrm{miss},2} \sim \int_{\theta_\mathrm{v,lim}}^{\pi/2} P(\theta_\mathrm{v})\,\mathrm{d}\theta_\mathrm{v}.
 \label{eq:Pmiss}
\end{equation}
Using the probability distribution $P(\theta_\mathrm{v})$ reconstructed\footnote{In low signal-to-noise ratio events such as GW190425, the posterior is well-approximated by the universal inclination angle distribution of GW-detected binaries of \citet{Schutz2011}. Indeed, using that distribution our result changes only by a few percent.} from the publicly available posterior samples \citep{Abbott2020_GW190425}, and assuming $\theta_\mathrm{v,lim}=0.2\,\mathrm{rad}$ (a representative value for an SGRB beaming angle, \citealt{Fong2015}), we obtain $P_{\mathrm{miss},2}\simeq 0.94$. 
This implies $P(J_2\,|\,f_\mathrm{j,GW},P_{\mathrm{miss},2}) \simeq 1 - 0.06 f_\mathrm{j,GW}$, which is the multiplicative correction to our result obtained by considering GW170817 only. The resulting posterior probability on $f_\mathrm{j,GW}$ is almost identical to the one obtained in \S\ref{sec:fjet_estimation}, as can be appreciated from Figure~\ref{fig:fjGW_multiple}, which is unsurprising given the shallow available constraints on GW190425.

Given this framework, a relevant question that can be addressed is how many secure identifications of a jet, and/or how many tight limits on its presence, are needed to constrain $f_\mathrm{j,GW}$ to a desired precision. In that respect, we must keep in mind the fact that, while the presence of a jet can be assessed with relative certainty in favourable cases such as GW170817, its absence is generally much more difficult to prove, especially due to the strong relativistic beaming of radiation, which typically makes the detection of a putative jet extremely difficult beyond some limiting viewing angle. In well-localised cases, constraints from deep late-time radio observations can be used to probe a larger range of viewing angles, but for the majority of events this will not be a feasible route due to the combination of large localisation error box and low expected brightness of the late-time radio afterglow \citep[e.g.][]{Hotokezaka2016,Dobie2021,Colombo2022}. This implies that it will be easier to constrain $f_\mathrm{j,GW}$ from below than from above. With these considerations in mind, we performed a simple experiment: we assumed a true value $f_\mathrm{j,GW,true}=0.5$ and constructed a sample of mock BNS events with jets assigned randomly with probability $f_\mathrm{j,GW,true}$. In order to represent in a simple way the jet detectability, for each event we randomly extracted a distance and an inclination from the universal joint distribution expected for GW-detected binaries \citep{Schutz2011}. We then assumed a distance-dependent limiting viewing angle 
\begin{equation}
\theta_\mathrm{v,lim}=\min\left[\frac{\pi}{2},\left(\frac{d_\mathrm{L}}{d_\mathrm{L,e}}\right)^{-2} \right]
\label{eq:thvlim}
\end{equation}
beyond which observations are unable to constrain the presence, or absence, of a jet. Here $d_\mathrm{L,e}$ is a parameter, which we set equal to the 25th percentile of the simulated distances, so that only for \resp{the closest} 25 percent of the events a jet can be excluded with 100\% confidence if not present. If, for the $i$-th binary, $\theta_\mathrm{v}<\theta_\mathrm{v,lim}$ and a jet was present, then we set $J_i=1$; in all other cases, we set $J_i=0$. For all events, we computed $P_{\mathrm{miss},i}$ from Equations \ref{eq:Pmiss} and \ref{eq:thvlim}, therefore assuming that no detection can be made beyond $\theta_\mathrm{v,lim}$, no matter how extensive the observations. 
\begin{figure}
 \centering\includegraphics[width=\columnwidth]{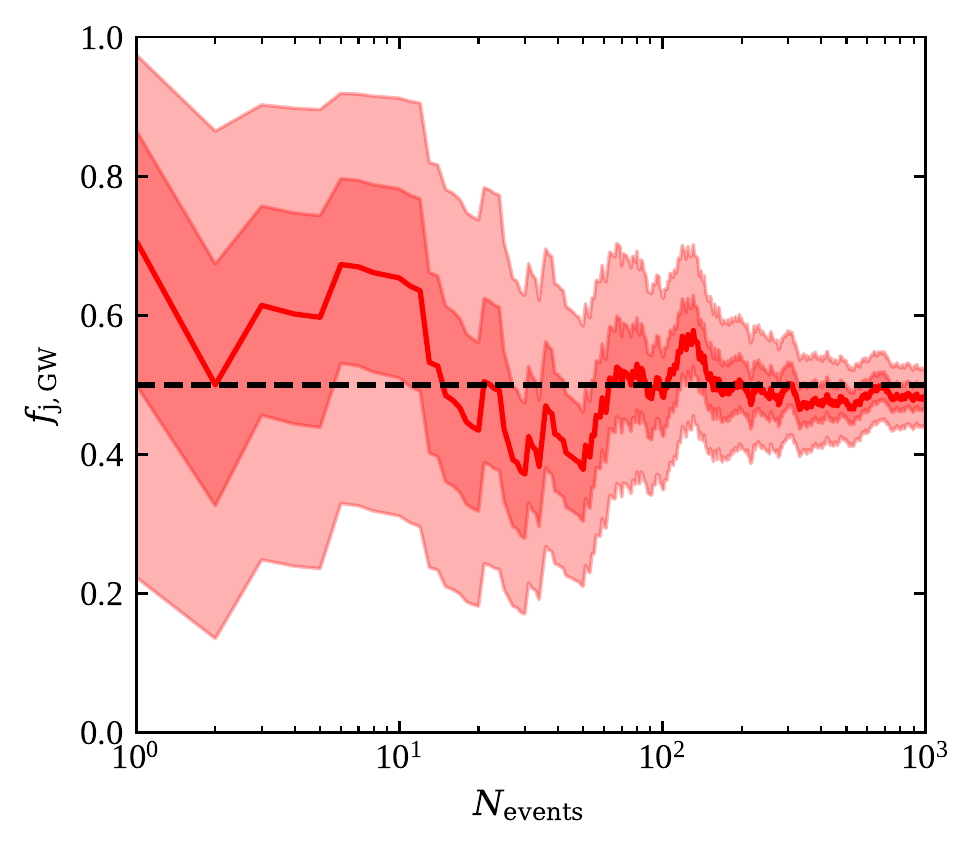}
 \caption{$f_\mathrm{j,GW}$ inference with multiple GW-detected BNS mergers. The coloured bands show the evolution of the posterior 90\% and 50\% credible range on $f_\mathrm{j,GW}$ after $N_\mathrm{events}$ simulated GW detections (see text), while the red solid line shows the median of the posterior. The horizontal black dashed line shows the true value.} 
 \label{fig:multi_event_fjGW_inference}
\end{figure}
Figure~\ref{fig:multi_event_fjGW_inference} shows how the posterior probability on $f_\mathrm{j,GW}$ evolves as we include an increasing number $N_\mathrm{events}$ of these simulated events in the inference. The result suggests that several tens of detections are needed to be able to constrain the jet \resp{incidence} to within 10-20\%, and hundreds to get to a few percent constraint.

\section{Local rate density of GRB170817A-like sources} \label{sec:local_sgrb_rate}

\begin{figure}
 \centering
 \includegraphics[width=\columnwidth]{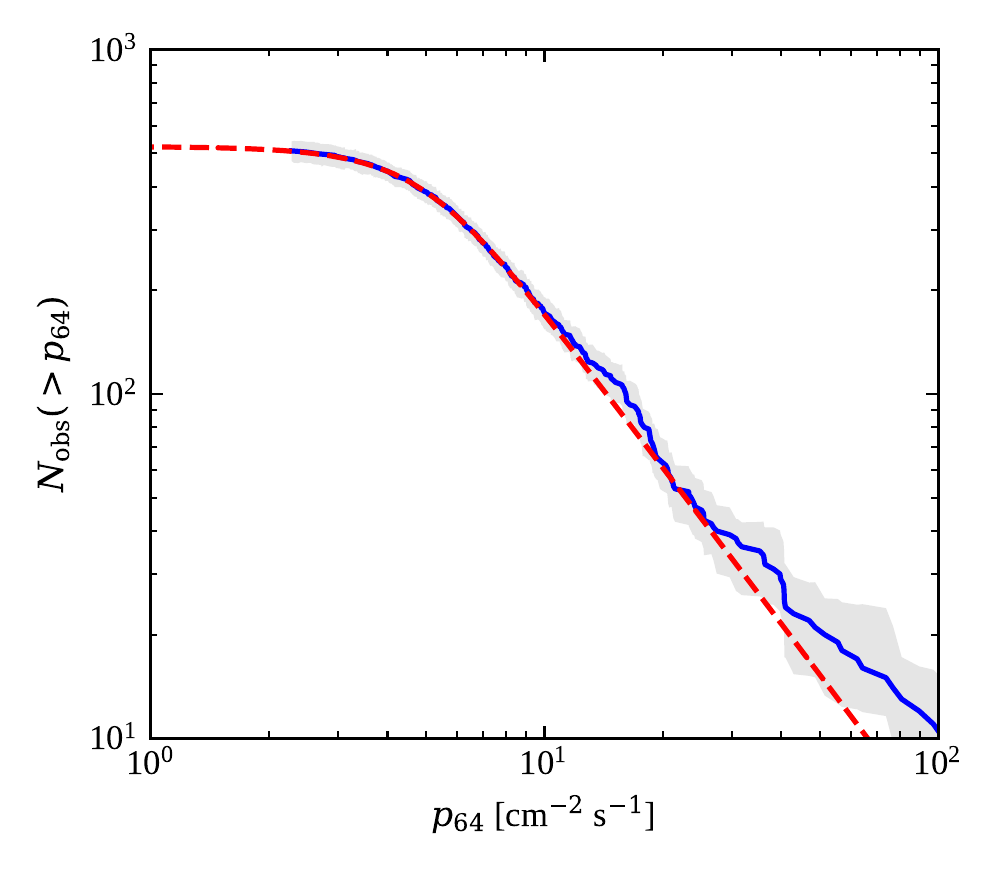}
 \caption{\small\textit{Fermi}/GBM observed inverse cumulative distribution of 64-ms-binned photon fluxes in the 10-1000 keV band (blue solid line, with the grey band showing the 90\% confidence band due to Poisson and measurement uncertainties), compared with our best-fitting model, Eq.~\ref{eq:Nobs_model}, with $\alpha=5.42$ and $p_{\frac{1}{2}}=2.25\,\mathrm{ph\,cm^{-2}\,s^{-1}}$ (red dashed line). The slight excess over the $p_{64}^{-3/2}$ expected trend at high fluxes is likely due to contamination of the sample by magnetar giant flares in local galaxies \citep{Burns2021}. }
 \label{fig:logNlogS}
\end{figure}

To estimate the local SGRB rate based on GRB170817A, we model the \textit{Fermi} Gamma-ray Burst Monitor (GBM) detection of GRB170817A-like sources (i.e.~sources with exactly the same luminosity and intrinsic gamma-ray spectrum as GRB170817A) as a Poisson process with an expected event number $\lambda = R_\mathrm{0}V_\mathrm{eff}T$ over a time $T=13\,\mathrm{yr}$ (the current duration of the GBM survey), where $V_\mathrm{eff}$ is the GBM effective sensitive volume. The posterior probability on $R_\mathrm{0}$, assuming a single detection, is

\begin{equation}
 P(R_0\,|\,1) \propto P(1\,|\,R_0) \pi(R_0)
 \label{eq:P(R0)}
\end{equation}
where $P(1\,|\,R_0)\propto R_0 \exp(-R_0 V_\mathrm{eff} T)$ and we adopt the Jeffreys prior $\pi(R_0)\propto R_0^{-1/2}$. We estimate the effective sensitive volume as
\begin{equation}
 V_\mathrm{eff} = \eta_\mathrm{GBM}\int_0^{\infty}\frac{1}{1+z}\frac{\mathrm{d}V}{\mathrm{d}z}P_\mathrm{det}(z)\mathrm{d}z
 \label{eq:VeffGBM}
\end{equation}
where $\eta_\mathrm{GBM}=0.59$ accounts for the GBM field of view and duty cycle \citep{Burns2016}, $\mathrm{d}V/\mathrm{d}z$ is the differential comoving volume \citep{Hogg1999}, \resp{the $(1+z)^{-1}$ factor accounts for cosmological time dilation} and $P_\mathrm{det}(z)$ represents the probability for GBM to detect a GRB170817A-like source at a redshift $z$ \resp{within its field of view}. In order to compute this quantity, we assume for simplicity that the GBM detection probability depends only on the peak photon flux $p_{64}$ of the GRB, measured with a 64 ms binning, in the 10-1000 keV band, so that $P_\mathrm{det}(z)=P_\mathrm{det}(p_{64}(z))$. We then make the ansatz

\begin{equation}
 P_\mathrm{det}(p_{64},\alpha,p_{\frac{1}{2}}) = \frac{1}{2}\left\lbrace1 + \tanh\left[\alpha\ln\left(\frac{p_{64}}{p_{\frac{1}{2}}}\right)\right]\right\rbrace,
 \label{eq:pdet_ansatz}
\end{equation}
which is a smooth function satisfying $P_\mathrm{det}=1/2$ when $p_\mathrm{64}= p_{\frac{1}{2}}$,  $P_\mathrm{det}\sim 1$ when $p_\mathrm{64}\gg p_{\frac{1}{2}}$ and $P_\mathrm{det}\sim 0$ when $p_\mathrm{64}\ll p_{\frac{1}{2}}$. The $\alpha$ parameter controls the sharpness of the transition from 0 to 1. To fix  $\alpha$ and $p_\mathrm{\frac{1}{2}}$, we proceed as follows. We assume the intrinsic inverse cumulative distribution of short GRB photon fluxes to follow $N_\mathrm{int}(>p_{64})\propto p_{64}^{-3/2}$, as expected for uniformly distributed sources in Euclidean space, in absence of evolution of the luminosity function with distance (which is reasonable as the typical redshifts are well below 1).  The observed distribution is then 
\begin{equation}
 N_\mathrm{obs}(>p_{64},\alpha,p_{\frac{1}{2}})\propto \int_{p_{64}}^\infty P_\mathrm{det}(p,\alpha,p_{\frac{1}{2}}) \frac{\mathrm{d}N_\mathrm{int}}{\mathrm{d}p}(p) \mathrm{d}p.
 \label{eq:Nobs_model}
\end{equation}
We fit the resulting distribution to that constructed with the actual \textit{Fermi}/GBM catalog data (restricted to events with $t_{90}\leq 2\,\mathrm{s}$) by minimizing the sum of the squares of the residuals, obtaining $\alpha=5.42$ and $p_{\frac{1}{2}}=2.25\,\mathrm{ph\,cm^{-2}\,s^{-1}}$. The observed and modelled distributions are shown in Figure~\ref{fig:logNlogS}.  

Given the GRB170817A measured \citep{Goldstein2017} peak photon flux $p_\mathrm{64}=3.7\,\mathrm{ph\,cm^{-2}\,s^{-1}}$, the best-fitting spectral model at the peak (second line of table 3 in \citealt{Goldstein2017}), the best-fitting host galaxy luminosity distance \citep{Hjorth2017} $d_\mathrm{L}=41\,\mathrm{Mpc}$, and assuming Planck cosmological parameters \citep{Planck2016}, we can compute the 64 ms photon flux of an identical source at any redshift, $p_{64}(z)$. This allows us to compute the integral in Eq.~\ref{eq:VeffGBM}, yielding $V_\mathrm{eff}=1.13\times 10^5\,\mathrm{Mpc^3}$. Plugging this in Eq.~\ref{eq:P(R0)}, we obtain the posterior probability distribution shown in Figure~\ref{fig:R0_R0BNS_comparison}, whose peak and 90\% credible interval are  $R_0=342_{-337}^{+1798}\,\mathrm{Gpc^{-3}\,yr^{-1}}$, in good agreement with the simpler estimate by \cite{DellaValle2018}. 

\section{Comparison with GWTC-3}\label{sec:GWTC3_comparison}
 
\begin{figure}
 \includegraphics[width=\columnwidth]{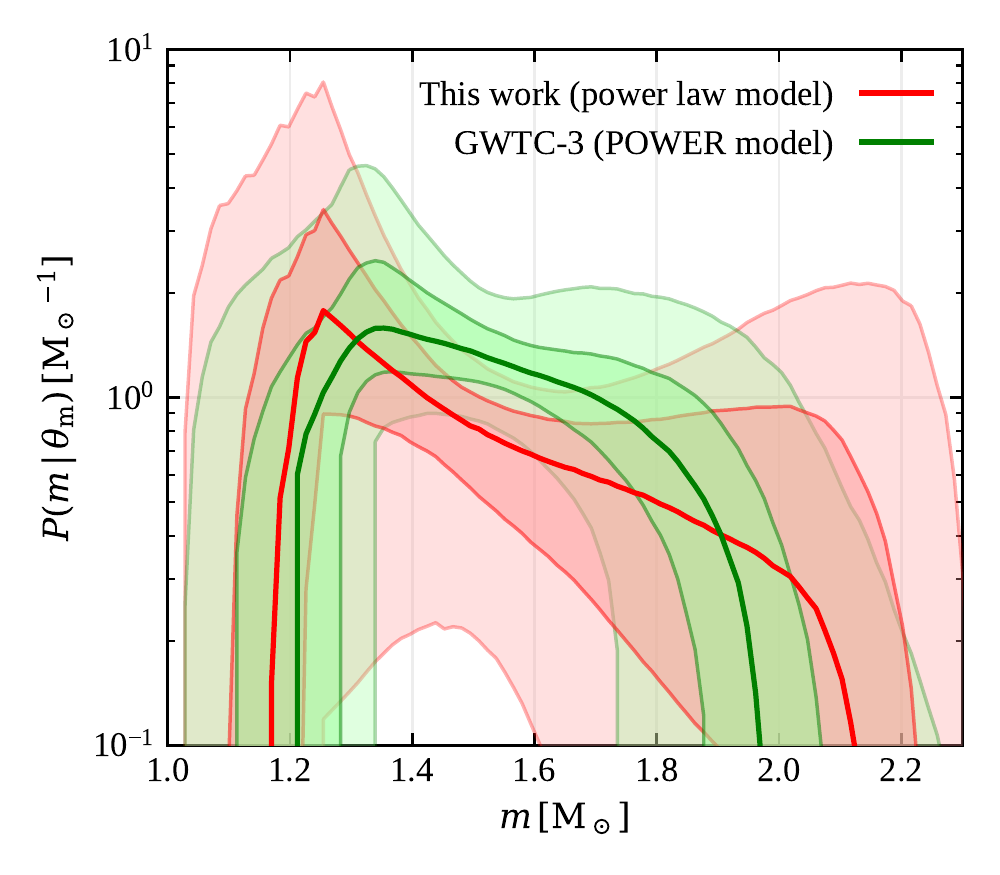}
 \caption{Comparison of our neutron star mass distribution posterior (power law model) with the GWTC-3 result from the population study described in \citealt{LVC2021_GWTC3pop} (their \textsc{POWER} model). Shaded areas show the 90\% and 50\% credible regions, while thick solid lines show the medians. Red is for our result (\S\ref{sec:mass_distrib_constraints}), blue is for the GWTC-3.}
 \label{fig:Plaw_GWTC3_comparison} 
\end{figure}
\begin{figure}
 \includegraphics[width=\columnwidth]{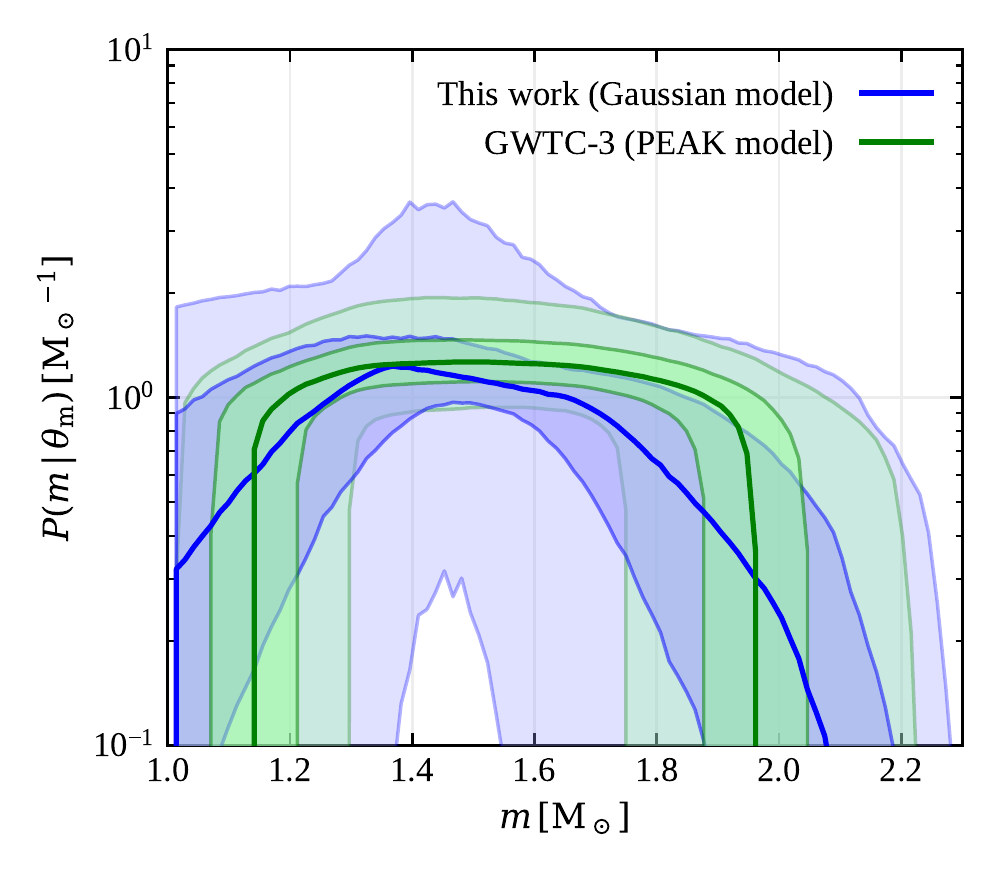}
 \caption{Same as Figure~\ref{fig:Plaw_GWTC3_comparison}, but for our Gaussian model and the GWTC-3 population study \textsc{PEAK} model.}
 \label{fig:Gauss_GWTC3_comparison}
\end{figure}

Figures~\ref{fig:Plaw_GWTC3_comparison} and \ref{fig:Gauss_GWTC3_comparison} show comparisons of our mass distribution posteriors with the results from \citet{LVC2021_GWTC3pop} based on GWTC-3 \citep{LVC2021_GWTC3} data. As a cautionary note, the definitions in our power law model match those of the \textsc{POWER} model from the GWTC-3 population study (Fig.~\ref{fig:Plaw_GWTC3_comparison}), but their \textsc{PEAK} model has the minimum mass cut $M_\mathrm{min}$ as an additional free parameter (while ours is fixed at $1\,\mathrm{M_\odot}$). Despite being significantly shallower, our constraints are in general agreement with those from the population study. 

\end{appendix}
\end{document}